\documentclass[twocolumn, longauth]{aa}
\usepackage{graphicx}
\usepackage{caption}
\usepackage{txfonts}
\usepackage{booktabs}
\usepackage{color}
\usepackage[dvipsnames]{xcolor}
\usepackage{natbib,twoopt}
\usepackage{lscape}
\usepackage{caption}
\usepackage{subcaption}
\usepackage[symbol]{footmisc}
\usepackage{tablefootnote}
\usepackage{multirow}
\usepackage[bookmarks=true, pdfnewwindow=true, colorlinks=true, allcolors=blue, breaklinks=true, final=true, citecolor=blue]{hyperref}
\usepackage{orcidlink}
\bibpunct{(}{)}{;}{a}{}{,}

\newcommand\farcd{\mbox{$.\!\!^{\circ}$}}%
\newcommand{\gpsc}{\mbox{g~cm$^{-2}$}}%
\newcommand{\massrate}{\mbox{$M_{\odot}\,\mathrm{yr}^{-1}$}}%
\newcommand{\kms}{\mbox{km\,s$^{-1}$}}%
\newcommand{\msun}{\mbox{$M_\odot$}}%
\newcommand{\lmsun}{\mbox{$L_\odot/M_\odot$}}%
\newcommand{\jybeam}{\mbox{Jy\,beam$^{-1}$}}%
\newcommand{\mjybeam}{\mbox{mJy\,beam$^{-1}$}}%

\newcommand{\um}{\mbox{$\mu$m}}%

\begin{document}

\title{Linear filament and nested cluster evolution tomography (LANCET)}

\subtitle{I. Capture the evolution of dense gas in 14-parsec filament G316.8}

\author{
Fengwei Xu\inst{\ref{kiaa}, \ref{mpia}} \thanks{fengweilookuper@gmail.com, fengwei@mpia.de} \orcidlink{0000-0001-5950-1932} \and 
Ke Wang\inst{\ref{kiaa}} \thanks{kwang.astro@pku.edu.cn} \orcidlink{0000-0002-7237-3856} \and
Nicola Schneider\inst{\ref{uzk}} \orcidlink{0000-0003-3485-6678} \and
Roberto Galván-Madrid\inst{\ref{irya}} \orcidlink{0000-0003-1480-4643} \and
Floris F.~S. van der Tak\inst{\ref{sron}} \orcidlink{0000-0002-8942-1594} \and
Adam Ginsburg\inst{\ref{uf}} \orcidlink{0000-0001-6431-9633} \and
Jonathan C. Tan\inst{\ref{uv}, \ref{dsee}} \orcidlink{0000-0002-3389-9142} \and
Hauyu Baobab Liu\inst{\ref{nsysu}, \ref{cag}} \orcidlink{0000-0003-2300-2626} \and
Qizhou Zhang\inst{\ref{cfa}} \orcidlink{0000-0003-2384-6589} \and
Wenyu Jiao\inst{\ref{shao}} \orcidlink{0000-0001-9822-7817}
Guido Garay\inst{\ref{dauc}, \ref{cassaca}} \orcidlink{0000-0003-1649-7958} \and
Sihan Jiao\inst{\ref{naoc},\ref{mpia}} \and
Keyun Su\inst{\ref{kiaa}} \and
Beth M. Jones\inst{\ref{uzk}} \orcidlink{0000-0002-0675-0078} \and
Lei Zhu\inst{\ref{cassaca}}
}

\institute{
\label{kiaa} Kavli Institute for Astronomy and Astrophysics, Peking University, Beijing 100871, People's Republic of China \and
\label{mpia} Max Planck Institute for Astronomy, K{\"o}nigstuhl 17, 69117 Heidelberg, Germany \and
\label{uzk} I. Physikalisches Institut, Universit{\"a}t zu K{\"o}ln, Z{\"u}lpicher Stra{\ss}e 77, 50937 Cologne, Germany \and
\label{irya} Instituto de Radioastronomía y Astrofísica, Universidad Nacional Autónoma de México, Morelia, Michoacán 58089, Mexico \and
\label{sron} SRON Netherlands Institute for Space Research \& Kapteyn Astronomical Institute, University of Groningen, 9747 AD Groningen, The Netherlands \and
\label{uf} Department of Astronomy, University of Florida, 211 Bryant Space Science Center, P.O. Box 112055, Gainesville, FL 32611-2055, USA \and
\label{uv} Department of Astronomy, University of Virginia, Charlottesville, VA 22904, USA \and
\label{dsee} Department of Space, Earth \& Environment, Chalmers University of Technology, 412 93 Gothenburg, Sweden \and
\label{nsysu} Department of Physics, National Sun Yat-Sen University, No. 70, Lien-Hai Road, Kaohsiung City 80424, Taiwan, ROC \and
\label{cag} Center of Astronomy and Gravitation, National Taiwan Normal University, Taipei 116, Taiwan \and
\label{cfa} Center for Astrophysics | Harvard \& Smithsonian, 60 Garden Street, Cambridge, MA 02138, USA \and
\label{shao} Shanghai Astronomical Observatory, Chinese Academy of Sciences, 80 Nandan Road, Shanghai 200030, People's Republic of China \and
\label{dauc} Departamento de Astronom\'ia, Universidad de Chile, Las Condes, 7591245 Santiago, Chile \and
\label{naoc} National Astronomical Observatories, Chinese Academy of Sciences, Beijing 100101, PR China \and
\label{cassaca} Chinese Academy of Sciences South America Center for Astronomy, National Astronomical Observatories, Chinese Academy of Sciences, Beijing, 100101, People's Republic of China
}

\date{Received ; accepted}

\titlerunning{LANCET. I. Dense gas evolution in G316.8}
\authorrunning{F. Xu et al.}

\abstract
{
A dynamic view of mass assembly is essential for understanding the formation of massive stars and clusters. However, interpreting evolutionary diagnostics from Galactic-wide surveys requires careful consideration of distance and environmental variations. The G316.8 filament provides an excellent controlled case: a 14-parsec, nearly linear structure comprising three contiguous subregions with comparable molecular gas reservoirs (each $\sim\!10{,}000$~\msun), yet spanning a clear evolutionary sequence from a northern infrared dark cloud (young) through a central massive young stellar object (intermediate), to a southern H{\sc ii} region (evolved). The {\it Linear filament and nested cluster evolution tomography} (LANCET) project mapped the entire G316.8 filament with the Atacama Compact Array (ACA) at 1.3~mm, achieving $6\arcsec$ (0.08~pc) resolution over $26.7$~arcmin$^2$ (17.1~pc$^2$). By combining ACA 7m data with \textit{Herschel} and APEX/ArT\'eMiS observations, we produced high-resolution temperature and column-density maps. We quantified subregional differences using (i) dense-fragment statistics, (ii) column-density probability distribution functions (N-PDFs), and (iii) the scale-dependent structural diagnostic, the $\Delta$-variance. From young to intermediate to evolved, the maximum fragment mass increases from 8 to 160 to 490~$M_\odot$, while the dense-gas mass fraction ($>0.5$~g~cm$^{-2}$) rises from $0.4\%$ to $2.3\%$ to $9.6\%$. Along this sequence, the N-PDF develops a slightly flatter primary power-law tail and an additional, steeper secondary tail; the $\Delta$-variance slope becomes progressively shallower. Across G316.8, the subregional differences consistently indicate a coherent evolutionary trend of massive star formation, in which gas is continuously assembled into sub-parsec dense structures. The forthcoming 12m array observations are about to extend this dynamic picture by resolving dense core formation and probing gas kinematics and magnetic fields.
}

\keywords{stars: formation – stars: massive – infrared: ISM – methods: observational – ISM: structure}

\maketitle

\section{Introduction}\label{sec:intro}

Massive stars dominate the radiative, mechanical, and chemical feedback in the Galactic ecosystem, yet a complete and self-consistent scenario for their formation remains unclear \citep[see reviews in][]{Zinnecker2007, Tan2014, Motte2018, Beuther2025}. Early theoretical studies primarily addressed the radiation pressure problem \citep{Wolfire1987} in the formation of individual massive stars, demonstrating that continued mass growth can be achieved through anisotropic, disk-mediated accretion \citep{Krumholz2009, Kuiper2011}. However, it has become increasingly evident from both observations and theory that massive stars rarely form in isolation. Instead, they predominantly emerge within embedded clusters that remain deeply enshrouded by their natal molecular clouds during the earliest evolutionary stages \citep{Lada2003}. Consequently, massive star formation must be considered in the broader context of cluster formation and evolution \citep{Motte2018, Beuther2025}. One of the key outstanding questions is how massive star clusters assemble their mass from their parental molecular cloud. Addressing this question is observationally challenging because the physical processes governing mass assembly, such as temperature, density, turbulence, and magnetic fields, are intrinsically time-dependent, whereas observations provide only instantaneous snapshots of these evolving systems \citep{Xu2023SDC335, Xu2024Assemble1}. Recovering the evolutionary sequence of massive cluster formation thus requires a dynamical perspective, achieved by systematically comparing sources across a wide range of evolutionary stages.

The bolometric luminosity over envelope mass $L_{\rm bol}/M_{\rm env}$, luminosity-to-mass ratio ($L/M$) in short, is widely used as an indicator of evolutionary stage in star-forming regions. In simplified models, massive star formation is treated as a scaled-up version of the classical inside-out collapse scenario developed for low-mass stars, in which envelope material accretes onto the central object while a fraction is removed through protostellar outflows. In this framework, the bolometric luminosity evolves as the sum of the time-dependent accretion luminosity, $L_{\rm acc}(t)$ and the intrinsic stellar luminosity, $L_\star(t)$, while the envelope mass $M_{\rm env}(t)$ decreases as accretion proceeds and feedback disperses the surrounding material. Such models predict evolutionary tracks in the $L_{\rm bol}{-}M_{\rm env}$ diagram and successfully reproduce the observed distributions of protostellar populations \citep[e.g.,][]{Saraceno1996, Molinari2008, DC2013}.

Because these models are formulated for individual protostellar objects, $L/M$ is, in principle, best measured within the immediate environment of massive stars, ideally on core scales ($\sim 0.01$~pc, or $\sim 2{,}000$~AU). However, achieving such spatial resolution remains challenging for distant ($\gtrsim 2$~kpc), massive star-forming regions. Using the point-process mapping (PPMAP) technique, \citet{Dellova2024} obtained unprecedented $2\farcs5$-resolution measurements toward the 15 ALMA-IMF massive clouds and demonstrated that $L/M$ measured on clump scales (a few times $0.1$ to $\sim$1~pc) remains sensitive to evolutionary stage. But this sensitivity is expected to be reduced when averaging over entire molecular clouds on parsec scales, where a significant fraction of the gas may not be directly involved in ongoing massive star formation. Recent numerical simulations further support the validity of clump-scale $L/M$ as a tracer of evolutionary progression \citep{Rosetta2}. 

Observational surveys of massive clumps spanning a wide range of $L/M$ values effectively sample different snapshots of the evolutionary sequence. Recent millimeter interferometric studies have revealed systematic variations in embedded massive protoclusters and their immediate gas environment. Using a sample of 13 massive clumps, \citet{Traficante2023} found that the number of dense cores marginally increases from $0.1$ to $100$~\lmsun\ (measured on $\sim\!0.1{-}0.3$~pc clump scales), while an expanded sample including data from \citet{Beuther2018} further showed a decreasing trend in core separations. Based on ALMA mosaics of 23 massive clumps ($\sim\!0.3{-}0.8$~pc) spanning the same $L/M$ range, \citet{Xu2024Assemble1} reported systematic core growth, enhanced core-clump mass correlations, decreasing core separations, and increasing mass segregation with increasing $L/M$. Large ALMA surveys have extended these results to statistically significant samples. The ALMAGAL program has observed more than 1000 massive clumps spanning from 0.04 to 500~\lmsun\ at scales of $\sim\!0.2{-}1.0$~pc, revealing a morphological evolution from compact, convex structures to more extended, complex shapes, together with confirmed core growth trends \citep{ALMAGAL1, ALMAGAL3}. The ALMA-IMF program targeted 15 massive protoclusters; it spans $7{-}160$~\lmsun\ at scales of $\sim\!0.2{-}0.6$~pc and identified 151 luminous protostellar objects, providing strong evidence for ongoing core mass growth during protostellar evolution \citep{Dellova2024, Motte2025}. Complementarily, using Atacama Compact Array (ACA) 7m observations of 178 massive clumps ($\sim$0.3--1~pc), \citet{Xu2024Quarks2} showed that the dense gas fraction increases from $\sim$1\% to $\sim$10\% over $0.05{-}500$~\lmsun, indicating progressive mass concentration as massive clumps evolve. 

However, interpreting evolutionary diagnostics from Galactic-wide surveys requires careful consideration of distance and environmental variations. First, the observed regions can span a wide range of distances (1--10~kpc), so either angular resolution should be carefully chosen: ALMA-IMF ($\sim$2,000~AU at 2--5~kpc) or ALMAGAL ($\sim$1,000~AU at 1--10~kpc). Second, these Galactic-wide regions are influenced by poorly constrained environmental effects and initial conditions. In the recent \textit{Rosetta Stone} project \citep{Rosetta1}, numerical simulations showed that in an individual massive clump ($\sim\!1$~pc scale) the correlation has a large scatter across different initial conditions, even though $L/M$ evolves monotonically from low to high \citep{Rosetta2, Rosetta3}. One solution is to select targets with subregions spanning different evolutionary stages, for example the W43 and W51 complexes in the ALMA-IMF survey.

Here we initiated the {\it Linear filament and nested cluster evolution tomography} (LANCET) project, which aims to conduct variable-controlled, comparative analyses of massive star-forming filaments. We selected elongated filaments containing at least two distinct evolutionary stages and observed them with high-resolution interferometric facilities under homogeneous instrumental setups. The underlying philosophy has precedent in {\it Multiwavelength line-Imaging survey of 70~$\mu$m dArk and bright clOuds} \citep[MIAO project;][]{Feng2020, Lin2025}, in which paired 70~$\mu$m-bright and -dark clumps within filamentary clouds were observed with the IRAM~30m telescope and compared in their chemical properties. Ideally, the rapid survey speed of state-of-the-art facilities such as ALMA enables an entire filament to be covered both within a compact time window to ensure uniform calibration and at high resolution and sensitivity to pinpoint toward dense cores and protostellar objects. With this design idea, the LANCET project is therefore uniquely suited to reveal dynamical and chemical differentiation across the lifetime of molecular clouds and therefore provide evolutionary views of massive star formation. 

In this paper, we first introduce our target G316.8 and its in situ evolutionary sequence in Sect.~\ref{sec:g316}. Second, we present the observations and data reduction in Sect.~\ref{sec:observe}. Then, we combine multiwavelength datasets to obtain multi-resolution dust temperature and column density images in Sect.~\ref{result:sed}. We perform three main analytical tools to characterize density structures at different evolutionary stages: structural hierarchy in Sect.~\ref{result:structure}, column density probability distribution functions (N-PDFs) in Sect.~\ref{result:npdf}, and $\Delta$-variance analyses in Sect.~\ref{result:delta}. Lastly, we put these results into the evolutionary context of massive star formation in Sect.~\ref{sec:evolution} and conclude in Sect.~\ref{sec:conclude}. 

\section{In situ evolutionary sequence in G316.8 filament} \label{sec:g316}

\begin{figure*}
\centering
\includegraphics[width=\linewidth]{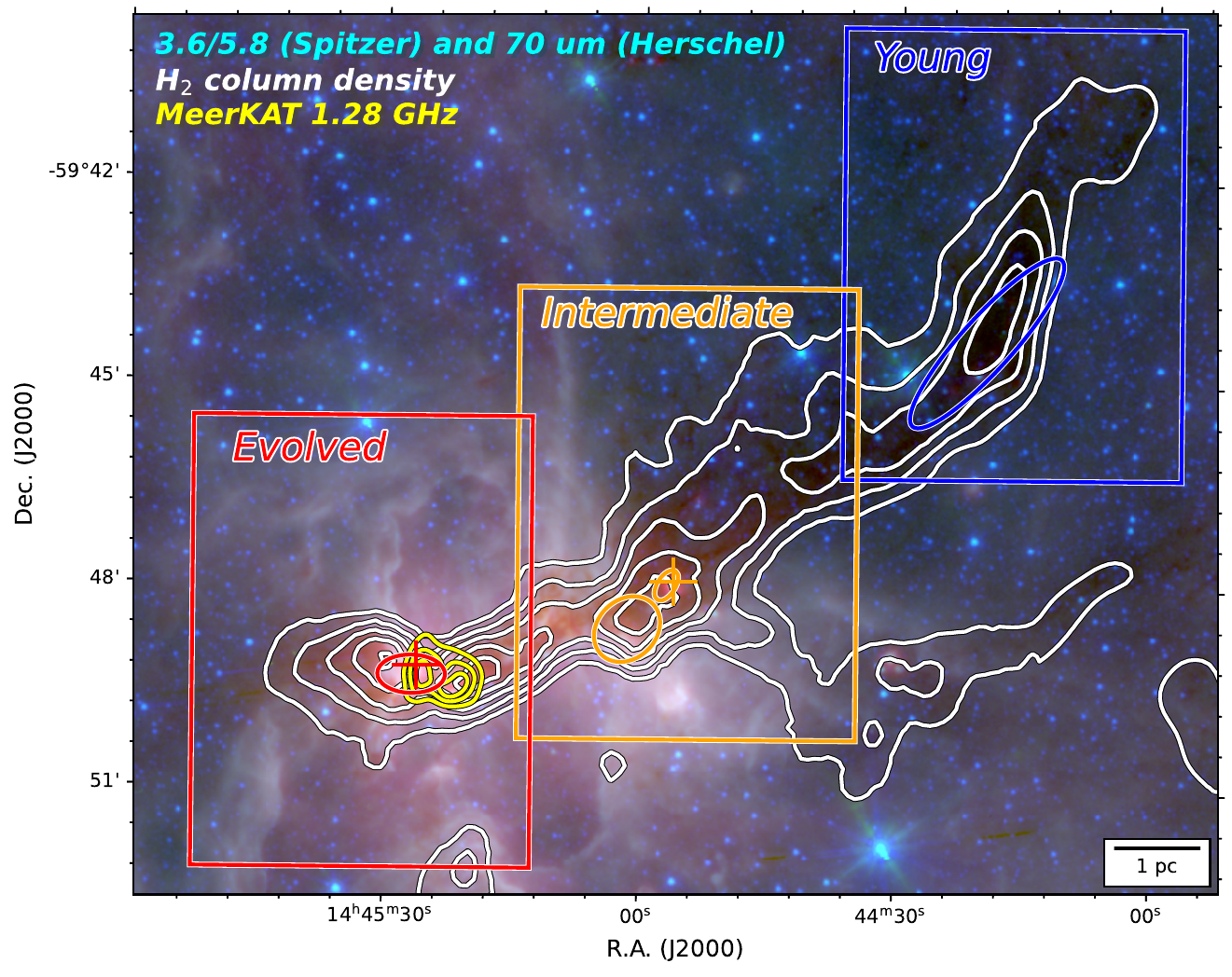}
\caption{Background color map composites of \textit{Spitzer} 3.6 and 5.8~$\mu$m and \textit{Herschel} 70~$\mu$m images. The overlaid white contours show the high-resolution H$_2$ column density map obtained from Sect.~\ref{result:sed} with levels of 3.0, 5.0, 8.0, 12.6, and $20.0 \times 10^{22}$~cm$^{-2}$. 
The blue, orange, and red boxes outline the young, intermediate, and evolved parts, respectively. Massive clumps embedded in each part are shown with color ellipses. The yellow contours show MeerKAT 1.28~GHz continuum emission at levels of 100, 180, and 260~\mjybeam\ (8\arcsec). The orange cross shows the G316.763$-$0.011 maser spot for OH and H$_2$O, while the red cross shows the G316.812$-$0.057 maser spot for OH, CH$_3$OH, and H$_2$O.}
\label{fig:overview}
\end{figure*}

\begin{table*}[!ht]
\caption{Physical properties of the three subregions}
\label{tab:globalprop}
\centering
\renewcommand{\arraystretch}{1.2}
\begin{tabular}{ccccccccc}
\hline
\hline
$\mathcal{R}$ & Size & $M_{\rm cloud}$ & $L_{\rm FIR}$ & $\langle N_{\rm H_2}\rangle$ & $\langle n_{\rm H_2}\rangle$ & $\langle T_{\rm dust} \rangle$ & $L_{\rm FIR}/M_{\rm cloud}$ & $L_{\rm bol}/M_{\rm clump}$ \\
 & [pc $\times$ pc] & [$10^4$~$M_\odot$] & [$10^4$~$L_\odot$] & [$10^{23}$~$\mathrm{cm}^{-2}$] & [$10^4$~$\mathrm{cm}^{-3}$] & [K] & [$L_\odot/M_\odot$] & [$L_\odot/M_\odot$] \\
\hline
Y & $4.4\times1.6$ & $1.1\pm0.4$ & $0.8 \pm 0.3$ & $1.6 \pm 0.5$ & $3.3 \pm 1.4$ & $14.4 \pm 0.1$ & $0.7 \pm 0.1$ & $0.7$ \\
I & $5.5\times1.3$ & $1.7 \pm 0.5$ & $6.1 \pm 2.4$    & $2.5 \pm 1.0$ & $6.5 \pm 2.5$ & $17.5 \pm 0.1$ & $3.6 \pm 0.4$ & $26$ \\
E & $2.9\times1.1$ & $0.8 \pm 0.3$ & $8.4 \pm 3.2$    & $2.5 \pm 1.0$ & $7.5 \pm 3.1$ & $21.3 \pm 0.1$ & $10.4 \pm 1.3 $ & $156$ \\
\hline
\end{tabular}
\tablefoot{Y: young; I: intermediate; E: evolved. Size: Length and width of filamentary clouds. $M_{\rm cloud}$: Subregion cloud mass. $L_{\rm FIR}$: Far-infrared luminosity. $\langle N_{\rm H_2}\rangle$: Mean column density of the region. $\langle n_{\rm H_2} \rangle$: Mean volume density assuming depth equal to width. $\langle T_{\rm dust} \rangle$: Mean dust temperature throughout the subregion. $L_{\rm FIR}/M_{\rm cloud}$: FIR luminosity-to-mass ratio calculated at cloud scale. $L_{\rm bol}/M_{\rm clump}$: Bolometric luminosity-to-mass ratio calculated at clump scale of $0.2{-}0.4$~pc.}
\end{table*}

On the near side of the Scutum-Centaurus Arm, the G316.8 filament lies at a heliocentric distance of $2.8\pm0.5$~kpc \citep{Reid2019}. It forms a continuous ridge extending over $\sim$14~pc and displays a remarkably clear internal evolutionary sequence. \citet{Watkins2019} quantified the impact of embedded O-type stars on the physical properties of the host cloud. They showed that, despite the presence of four O-type stars, the G316.8 ridge remains globally gravitationally bound with a virial parameter $\alpha_{\rm vir} \le 2$ across most of the filament, as expected for dense filamentary structures. With the high angular resolution achieved in this work, these in situ evolutionary contrasts make G316.8 an ideal laboratory for characterizing the structural evolution of massive star-forming regions and for investigating how gas is dynamically assembled during massive star cluster formation.

The northern portion of G316.8, outlined by the blue rectangle in Fig.~\ref{fig:overview}, is classified as a \textit{Spitzer}-dark cloud \citep{Peretto2009}. It contains a massive clump, AGAL316.719$+$00.076, identified in the APEX Telescope Large Area Survey of the Galaxy \citep[ATLASGAL;][]{Contreras2013, Urquhart2014}. It remains dark from 3.6 to $70~\um$, indicating cold ($\sim$10~K) and dense ($A_V > 30$~mag) gas with little evidence for stellar feedback \citep[e.g.,][]{Rathborne2006, Butler2009, Kainulainen2011}. Due to its lack of detectable mid-infrared emission, we assume that its bolometric luminosity is dominated by far-infrared (FIR) luminosity, with $L_{\rm bol} \simeq L_{\rm FIR} \sim 0.8 \times 10^4$~$L_\odot$. The region exhibits a mean surface density of $\gtrsim 0.1$~\gpsc\ and contains a mass reservoir of $\sim\, 1.1 \times 10^4$~$M_\odot$ within an effective radius of $\sim 2$~pc, exceeding commonly adopted thresholds for high-mass star formation \citep{Kauffmann2010, Urquhart2014, He2015}. Within the ATLASGAL footprint, the low luminosity-to-mass ratio of $\sim\! 0.7$~\lmsun\ is ten times lower than the lowest value among the ALMA-IMF young group. However, it is comparable with the 70~\um-dark sample $\sim\! 0.1{-}1~\lmsun$ \citep[e.g.,][]{Sanhueza2019, Morii2024}. Its early nature is further supported by the absence of MeerKAT 1.28~GHz radio continuum emission \citep{Goedhart2024}. We classify it as a young cloud.

The central intermediate part is a trident hub–filament system \citep[HFS;][]{Myers2009}. The major filament follows the G316.8 northwest-southeast spine, while the other extends toward the southwest. At the convergence, there are two massive clumps: AGAL316.764$-$00.012 and AGAL316.768$-$00.026. The first one hosts a 70~$\mu$m point source \citep{Elia2017HiGAL} as well as hydroxyl (OH) and water (H$_2$O) maser spots \citep{Breen2010}, as marked by the orange cross in Fig.~\ref{fig:overview}. Its bolometric luminosity of $\sim3\times10^{3}\,L_\odot$ \citep{Urquhart2022} corresponds to a single B1--B2 type zero-age main-sequence star ($M_\star\simeq10\,\msun$) or, equivalently, to an $\sim8\,\msun$ massive young stellar object (MYSO) accreting at a rate of $\dot{M}\sim10^{-4}~\massrate$. However, no associated ionized gas is detected at centimeter wavelengths, indicating that the region has not yet evolved to the stage of developing an ultracompact (UC) H{\sc ii} region. From the integrated energy spectrum from 3 to 870~\um, the clump-scale $L_{\rm bol}/M_{\rm clump}$ is estimated to be $26~\lmsun$ \citep{Urquhart2022}, consistent with the median value of ALMA-IMF intermediate group.

The G316.8 southern end hosts a well-studied IRAS source I14416$-$5937. It is rich in signposts of high-mass star formation, including masers, UC H{\sc ii} regions, and a compact X-ray source \citep[and references therein]{Watkins2019}. In the bottom-right panel of Fig.~\ref{fig:overview}, there are two 1.28~GHz emission peaks, both spatially consistent with the 24~GHz continuum from Australia Telescope Compact Array (ATCA) observations at 8-10\arcsec\ resolution. They trace free-free emission by ionized gas with emission measure (EM) of approximately $5{-}7\times10^6$~pc\,cm$^6$ \citep{Longmore2007}. At the red cross, H$_2$O, OH, and CH$_3$OH masers are found \citep{Breen2010}, which are associated with a massive clump AGAL316.812$-$0.057 identified in the ATLASGAL survey. \citet{Shaver1981} argued that a single O6 star could power the observed 1.4~GHz free–free emission. Radiative-transfer modeling that combines near-infrared, FIR, and radio data indicates two embedded sources (IRAS~14416$-$5937~A and B) requiring O-type stars of 45~\msun\ and 25~\msun\ to reproduce the observed luminosities \citep{Vig2007}. 2MASS color–magnitude and color–color analyses further reveal a large population of B0 or earlier-type stars in the field. Kinematic studies of the ionized gas using radio recombination lines show a pronounced velocity gradient, plausibly tracing the remnant angular momentum of the parental cloud \citep{Longmore2009, Dalgleish2018}. Above all, these signatures confirm that the southern region harbors a forming OB cluster and that, more importantly, it has already begun to inject strong stellar feedback into its surroundings. The clump-scale $L_{\rm bol}/M_{\rm clump}$ is measured as $156~\lmsun$ \citep{Urquhart2022} -- close to the value $160~\lmsun$ of G333.60, the most luminous one in the ALMA-IMF sample. We therefore classify the southern part as the evolved stage. 

\section{Observations} \label{sec:observe}

\subsection{ACA 7m observations}
\label{observe:alma}

\begin{figure}[!ht]
\centering
\includegraphics[width=\linewidth]{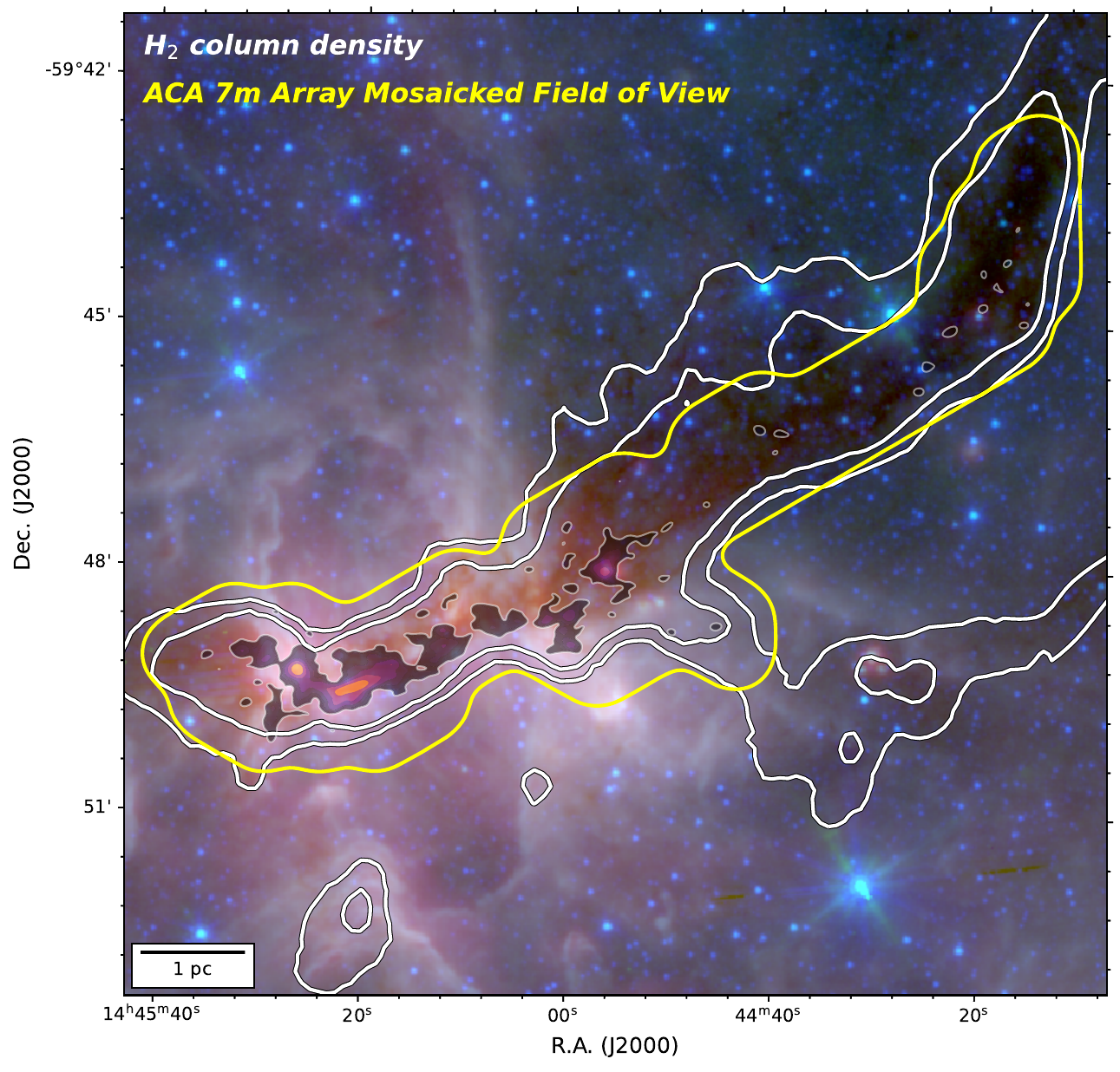}
\caption{ACA 7m continuum emission at 1.3~mm overlaid on the composite \textit{Spitzer} 3.6~$\mu$m and 5.8~$\mu$m emission and \textit{Herschel} 70~$\mu$m emission. The 1.3~mm map is shown with filled contours at levels of 10, 20, 40, 80, 160, and 320~\mjybeam. The mosaicked field of view is outlined by primary beam response (yellow contour at 0.2). The white contours define the dense region at levels of $3.0$ and $5.0 \times 10^{22}$~cm$^{-2}$.} 
\label{fig:observe}
\end{figure}

The G316.8 filament was observed by the ACA 7m arrays in the ALMA project 2016.1.00909.S (PI: Ke Wang). 118 pointings of 7m array were used to cover the whole filament. The ACA 7m data were acquired during the ALMA Cycle-4, with eight observing dates spanning from December 4, 2016, to April 28, 2017. The baselines range from 8.9 meters to 48.9 meters. The total on-source integration time of the ACA 7m array is $\sim\!5.4$ hours. The spectral tunings were set both to achieve a wide bandwidth for continuum aggregation and to cover several lines of interest in six narrow windows. Two wide spectral windows, namely SPWs 1--2, were centered at 216.732~GHz and 230.940~GHz, respectively, each with a 1.875~GHz bandwidth across 3840 channels. The channel width is $\Delta V \sim\!488$~kHz, i.e., 0.63--0.67~\kms. Six high-resolution spectral windows, namely SPWs 3--8, were centered at 218.248, 218.484, 219.587, 230.565, 231.249, and 231.350~GHz. The channel width is $\Delta V \sim\!61$~kHz, corresponding to 0.79--0.84~\kms. They were respectively aimed at H$_2$CO ($3_{03}$-$2_{02}$), H$_2$CO ($3_{22}$-$2_{21}$), C$^{18}$O (2-1), CO (2-1), $^{13}$CS (5-4), and N$_2$D$^+$ (3-2). Among the narrow windows, SPWs 4 and 6 have bandwidth of 117~MHz, corresponding to 152--160~\kms. The others have bandwidth of 59~MHz or 76--81~\kms. This paper only discusses the ACA 7m continuum data; the spectral line data will be presented in future papers. 

\subsection{Data reduction}
\label{observe:reduce} 

The ACA 7m array data were routinely calibrated using the ALMA pipeline of the Common Astronomy Software Applications \citep[CASA;][]{CASA2022}, versions 5.4.0-70. The atmosphere and pointing calibrators are J1256$-$0547 and J1326$-$5256. The bandpass calibrator is J1256$-$0547. The flux calibrator is Ganymede. The phase calibrator is J1326$-$5256. The flux calibration error is typically $\sim5{-}10\%$ for Band~6 

The line emission channels of the two wide bands were manually flagged and then averaged every 512 channels (width of 250~MHz) to produce channel-binned continuum visibilities, which were used for continuum imaging. The line-free channels were selected and then fit in the Fourier space with \texttt{uvcontsub}, using a polynomial order of 1, to obtain continuum-subtracted visibilities for line cube imaging. 

The continuum and line cube imaging processes were performed using the \texttt{tclean} task in CASA 6.5.6, with a \texttt{briggs} robust weighting of 0.5. Automatic masking was adopted using the \texttt{auto-multithresh} algorithm with input parameters as recommended by the official guides \footnote{\url{https://casaguides.nrao.edu/index.php?title=Automasking_Guide}}. A total of 118 pointings were mosaicked with an image size of $750\times576$\,pixel$^2$ and a cell size of 1\arcsec. The representative frequency of the continuum image is 223.8~GHz, i.e., 1.3~mm. The continuum image beam is 7\farcs4$\times$5\farcs3 with position angles (PAs) of $-$85\farcd4. The maximum recoverable scale (MRS) is about 28\farcs0. 

We performed self-calibration to correct the phase errors. We adopted the same workflow as \citet{Ginsburg2018} where phase solutions with low signal-to-noise were dropped 
\footnote{\url{https://github.com/keflavich/SgrB2_ALMA_3mm_Mosaic}}. After one round of self-calibration, some artifacts surrounding the bright source were eliminated, and the noise rms of the continuum image improved from 4 to about 2~\mjybeam. This corresponds to a mass sensitivity of 0.3~$M_\odot$ if assuming a temperature of 20~K, dust opacity of $0.9$~cm$^2$~g$^{-1}$ \citep[see Fig.~D.1 in][]{Xu2025}, and gas-to-dust ratio of 100. The final continuum image is shown by color-filled contours in Fig.~\ref{fig:observe}.

\section{Results} \label{sec:result}

\subsection{Multi-resolution temperature and column density images} \label{result:sed}

\begin{figure*}[!ht]
\centering
\includegraphics[width=0.9\linewidth]{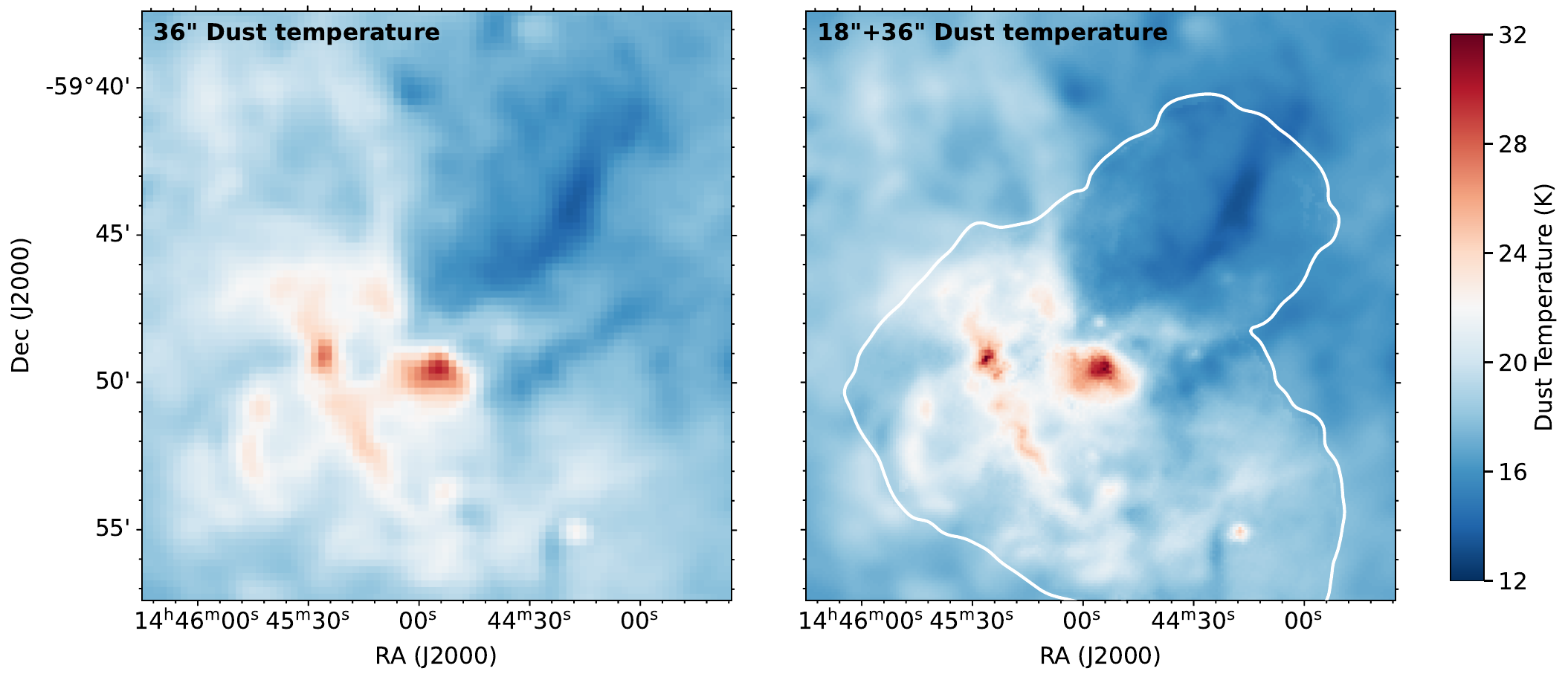}
\includegraphics[width=0.9\linewidth]{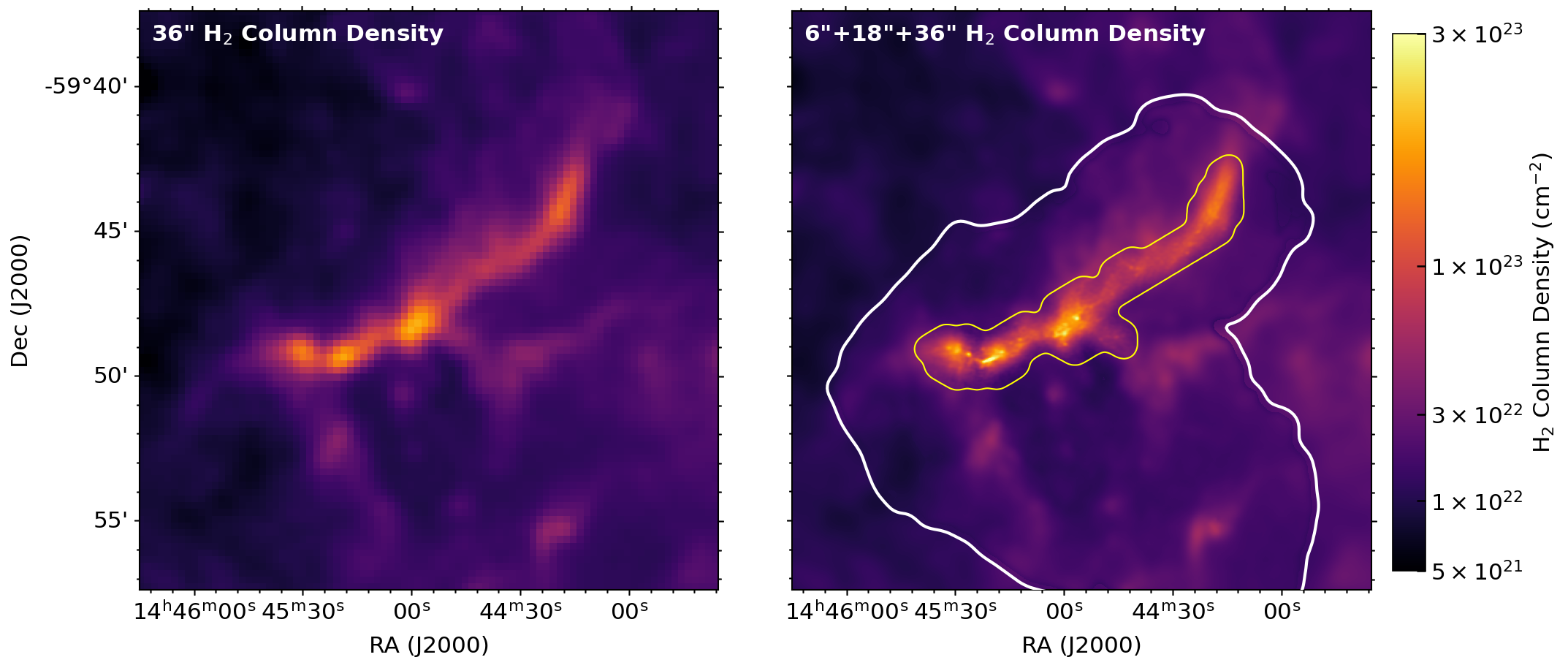}
\caption{Comparison between low-resolution and multi-resolution maps of dust temperature (upper) and H$_2$ column density (lower). The white and yellow polygons indicate the fields of view of the APEX and our ACA 7m observations, respectively. For the temperature image, the resolution is 18\farcs2 inside and 36\farcs3 outside the white closed polygon. For the column density image, the resolution outside the yellow closed polygon is the same as the temperature map but as high as $6\arcsec$ within the yellow closed polygon.}
\label{fig:multires}
\end{figure*}

The dust temperature and H$_2$ column density images were obtained from spectral energy distribution (SED) fitting. The datasets include \textit{Herschel} infrared Galactic Plane Survey \citep[Hi-GAL;][]{Molinari2010,Molinari2016} at PACS 160~$\mu$m, and SPIRE 250, 350, and 500~$\mu$m with effective full width at half maximum (FWHM) resolutions of 13\farcs5, 18\farcs2, 24\farcs9, and 36\farcs3. The PACS photometry has a flux calibration uncertainty of 7\%, while the SPIRE photometry has an uncertainty of 5.5\%. The 70~$\mu$m data were not used in our fitting due to their potential contamination by hot dust components from protostars. We also utilized the 350 and 450~$\mu$m images from the APEX/ArT{\'e}MiS CAFFEINE survey with 8\arcsec\ and 10\arcsec\ resolution, respectively \citep{Mattern2024}. The absolute flux calibration uncertainty is about 10\% but was corrected using the \textit{Planck} maps \citep{Planck2011}. The published data\footnote{\url{https://sites.google.com/view/artemis-apex-caffeine/image-products}} were then combined with \textit{Herschel} data to recover large-scale flux. The rms noise is about 0.5~\jybeam at 350~$\mu$m and about 0.4~\jybeam at 450~$\mu$m. 

Canonical and conservative SED fitting smooths all images to the coarsest resolution of 36\farcs3 and regrids them to the same pixel size of 14\farcs0. Under the assumption of single-component graybody emission and the dust opacity model for diffuse interstellar clouds \citep[H83;][]{Hildebrand1983}, pixel-wise temperature and column density were fit (for further details, see Appendix~\ref{app:sedfitting}). Temperature uncertainties range from 0.1--0.8~K, while column density uncertainties are about 10\% per pixel. The 36\farcs3-resolution maps are shown in the left column of Fig.~\ref{fig:multires}. While sufficient for nearby star-forming regions, the method cannot pinpoint dense cores in distant regions \citep[e.g.,][]{Jiao2025b}.

To this end, we did not simply smooth all images to the coarsest resolution but instead made full use of their intrinsic resolution information to construct so-called multi-resolution images. Medium-resolution SED fitting was performed within the APEX/ArT{\'e}MiS footprint, delineated by the white polygons in Fig.~\ref{fig:multires}. We used the \textit{Herschel} 160 and 250~$\mu$m bands together with the APEX/ArT{\'e}MiS 350 and 450~$\mu$m data to produce $18\arcsec$-resolution images in the canonical way. The key improvement is that the two APEX/ArT{\'e}MiS bands provide crucial additional data points, yielding more reliable SED fits at $18\arcsec$ resolution. We then merged the low- and medium-resolution solutions in the image domain via edge fusion (see Appendix~\ref{app:comb:xy}) to obtain combined $18\arcsec$+$36\arcsec$ temperature and column-density maps. As indicated by the black solid cross in Fig.~\ref{fig:sedpixel}, the $18\arcsec$-resolution maps were further used for pixel-wise extrapolation of the 1.3~mm emission. The resulting $18\arcsec$-resolution 1.3~mm map was then combined with the ACA 7m continuum image at $6\arcsec$ resolution in the $uv$ domain, as described in Appendix~\ref{app:comb:uv}. The linear-combination algorithm follows a similar principle to that adopted by the ALMA-IMF team using \texttt{feather} \citep{Diaz2023} but employs a modified weighting function to achieve a near-Gaussian synthesized beam response, as demonstrated in \citet{Jiao2022}.

\begin{figure*}[!ht]
    \centering
    \includegraphics[width=\linewidth]{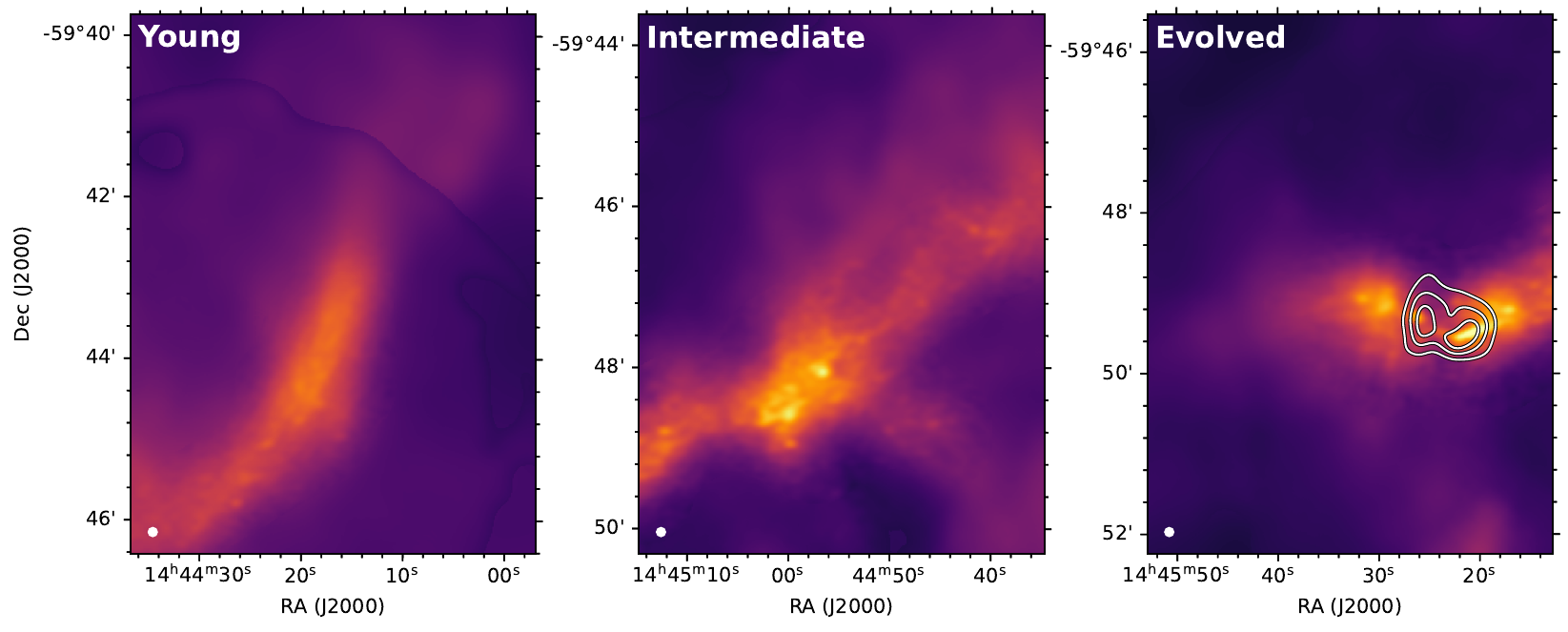}
    \caption{High-resolution column density images zoomed-in toward the three subregions: young (left), intermediate (middle), and evolved (right). The $6\arcsec$ resolution is shown in the bottom left corner. The white contours show MeerKAT 1.28~GHz continuum emission at levels of 100, 180, and 260~\mjybeam.}
    \label{fig:hires_NH2}
\end{figure*}

With a medium-resolution ($18\arcsec$) dust temperature ($T^{\rm midres}_{\rm dust}$), we converted the combined 1.3~mm image $I^{\rm comb}_{\rm 1.3~mm}$ into a high-resolution column density map by
\begin{equation} \label{eq:monoband}
    N^{\rm hires}_{\rm H_2} = \frac{-\ln\left[1 - I^{\rm comb}_{\rm 1.3~mm} / B_{\rm 1.3~mm} (T^{\rm midres}_{\rm dust}) \right]}{\mu m_{\rm H} \kappa_{\rm 1.3~mm}}.
\end{equation}
The assumption here is that dust temperature changes slowly and that the emission at this wavelength originates purely from thermal dust emission. The obtained high-resolution column density map $N^{\rm hires}_{\rm H_2}$ has a resolution of 6\farcs2, or 0.08~pc in physical scale. Similarly, we merged it with the 18\arcsec$+$36\arcsec\ map in the image domain to produce the multi-resolution (6\arcsec$+$18\arcsec$+$36\arcsec) column density map following the procedure in Appendix~\ref{app:comb:xy}. The multi-resolution column density map is displayed in bottom right of Fig.~\ref{fig:multires}. 
As shown in Fig.~\ref{fig:hires_NH2}, three subregions at different evolutionary stages are zoomed in. A distinct morphological difference exists in the three subregions: the young part globally shows a flat density profile with faint, isolated speckles; while the intermediate and evolved parts show more substructures, bright, clustered sources, and higher density contrast.

We note two caveats for the column density associated with the assumption above. The biases introduced by the slowly varying temperature assumption essentially smooth the temperature contrast while enhancing the column density contrast. If the temperature is underestimated, the column density is overestimated (and vice versa). Based on the PPMAP method, the ALMA-IMF community found global dust temperatures as high as 40~K for evolved sources \citep[e.g.,][]{Dellova2024}. While in our case, the dust temperature reaches 33~K. To give an estimate, at 1.3~mm, adopting the assumed 33~K instead of the true 40~K overestimates the inferred column density by $N_{\rm assu}/N_{\rm true}\sim B_\nu(40\,\mathrm{K})/B_\nu(33\,\mathrm{K})\simeq1.25$, i.e., $\sim25\%$ in the optically thin limit. The other caveat arises from free-free contamination. Using archived the ATCA 24~GHz data from \citet{Longmore2007}, we found that the free-free contamination is at most 0.08~\jybeam\ at 1.3~mm. But the corresponding 1.3~mm emission is as bright as 0.53~\jybeam, so the free-free contamination is at most 14\% in an 8\arcsec\ beam. Further millimeter radio recombination lines such as H30$\alpha$ could be used for free-free subtraction \citep[e.g.,][]{GM2024}. 

\subsection{Hierarchical structures and physical properties} \label{result:structure}

Based on the multi-resolution images, we identified hierarchical structures and characterized their physical properties. We first characterized the global physical properties of the three subregions in Fig.~\ref{fig:overview}. Using the 18\arcsec-resolution column density images in Sect.~\ref{result:sed}, we defined the region of interest (\texttt{roi}) as those with column density higher than $3\times10^{22}$~cm$^{-2}$ or $\sim0.14$~g~cm$^{-2}$. Within the \texttt{roi} mask, the length $L$ and width $w$ are defined as the major and minor axes of the ellipse with the same normalized second moments of intensity. The cloud mass $M_{\rm cloud}$ and FIR luminosity $L_{\rm FIR}$ as well as uncertainty were then computed as the summation of all enclosed pixels within the \texttt{roi} mask: 
\begin{equation} \label{eq:enc}
    X_{\rm enc} = \mathcal{A} \sum^{\in \rm roi}_{\rm pixel} X_i \quad
    \sigma_{X} = \left(\sum^{\in \rm roi}_{\rm pixel} \sigma_{X,i}^2\right)^{1/2},
\end{equation}
where $X_i$ and $\sigma_{X,i}$ are parameter and uncertainty maps, and $\mathcal{A}$ is the physical area of one pixel. The mean volume density is defined as $\langle{n_{\rm H_2}}\rangle = \langle{N_{\rm H_2}}\rangle/w$,
where $\langle{N_{\rm H_2}}\rangle$ is the mean value of column density and $\langle T_{\rm dust}\rangle$ is the mean dust temperature within the \texttt{roi} mask. To gauge the evolutionary stages, the bolometric luminosity-to-mass ratio at clump scale ($L_{\rm bol}/M_{\rm clump}$) was retrieved from \citet{Urquhart2022}. All these physical properties are listed in Table~\ref{tab:globalprop}. 

\begin{figure*}[!ht]
\centering
\includegraphics[width=0.48\linewidth]{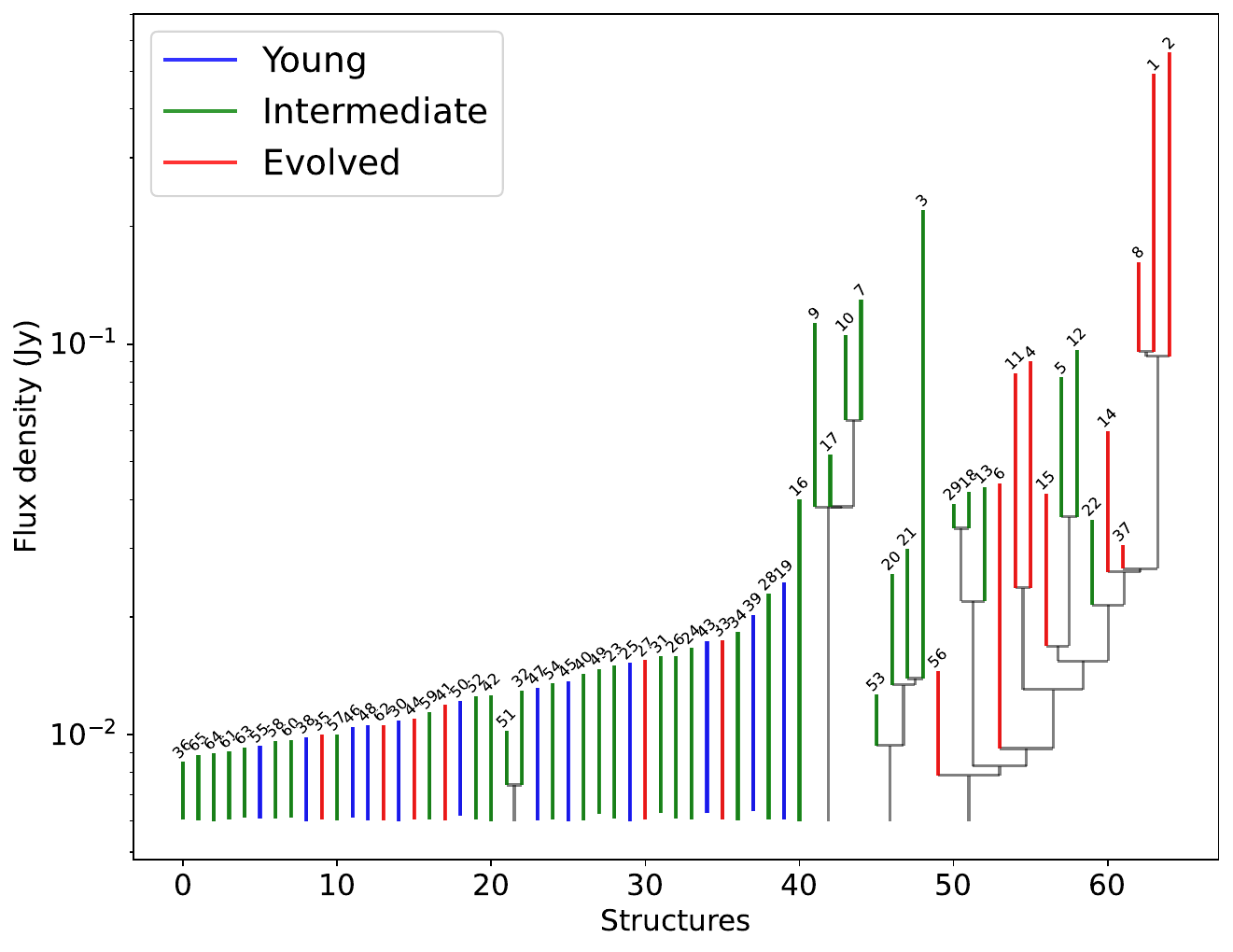}
\includegraphics[width=0.5\linewidth]{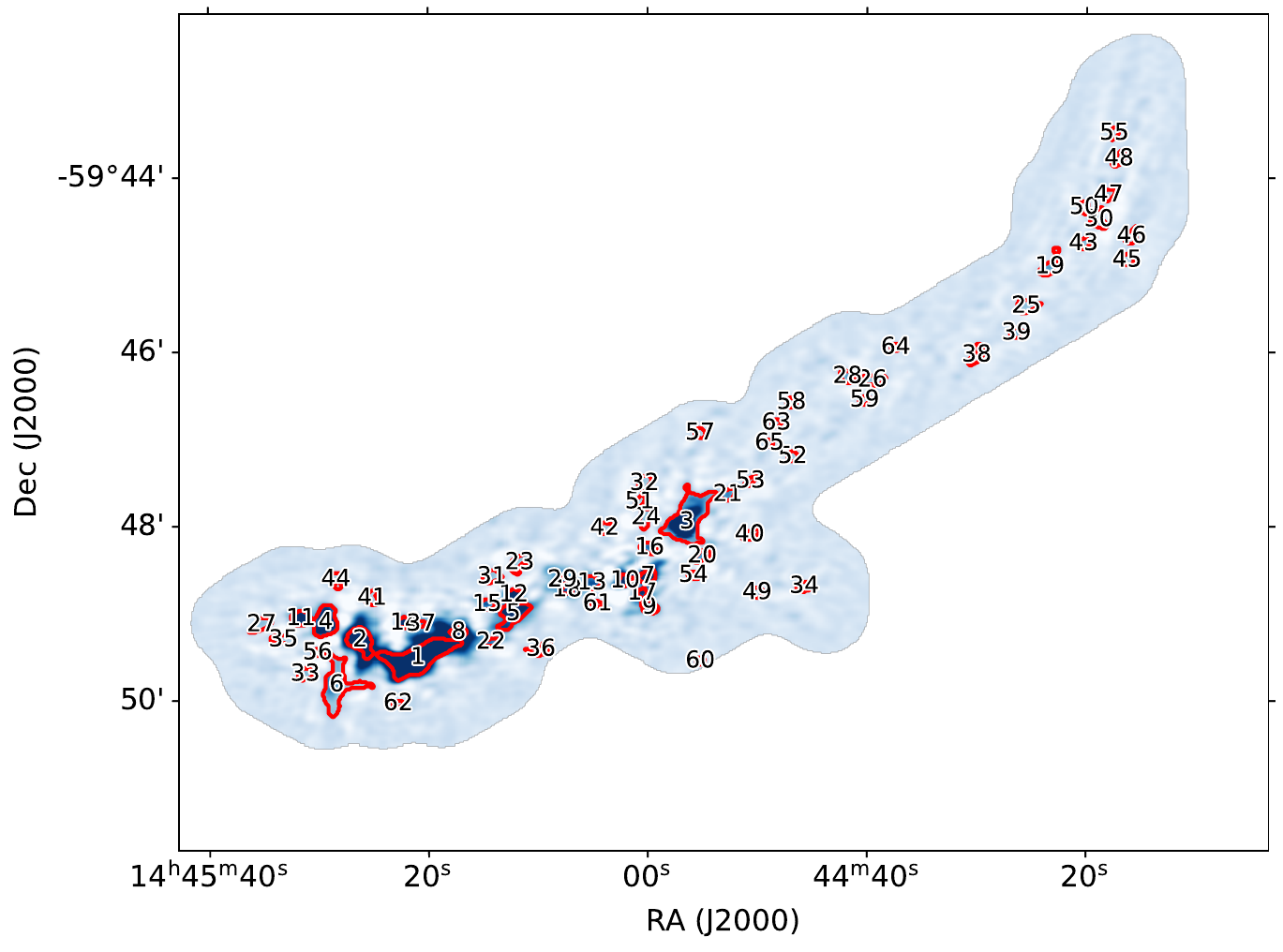}
\caption{Result of source extraction by \textit{astrodendro}. \textit{Left}: Dendrogram plot highlighting the leaf structures and their locations. \textit{Right}: ID numbers and region masks (red contours) overlaid on the 1.3~mm continuum map. 
}
\label{fig:astrodendro}
\end{figure*}

High-resolution data enable new discoveries of numerous dense structures at smaller scales. From visual inspection of Fig.~\ref{fig:multires} and \ref{fig:hires_NH2}, the structures show distinct variations in brightness and compactness from the northern to the southern part. To better characterize these differences, we used the \textit{astrodendro} algorithm \citep{Rosolowsky2008} to extract dense structures in the 1.3~mm continuum map. We adopted a intensity threshold of $3\sigma$, a step of $1\sigma$, and a minimum number of pixels equal to those contained in half of each synthesized beam. In Fig.~\ref{fig:astrodendro}, the hierarchical tree-like structures are shown, with leaves overlaid on the continuum map. A total of 65 leaves were identified and extracted. They are listed in descending order of flux density in Table~\ref{tab:catalog}. From the dendrogram, the young part contains only simple leaf structures, while the intermediate and evolved parts have complex tree structures.

\begin{table*}[!ht]
\caption{Catalog of \textit{astrodendro} leaves}
\label{tab:catalog}
\centering
\begin{tabular}{lcccccccccccccc}
\hline
\hline
ID & RA & Dec & $F_\nu$ & $\theta_{\rm maj}$ & $\theta_{\rm min}$ & PA & $R_\theta$ & $R_p$ & $R_{\rm \theta, dec}$ & $R_{p,\rm dec}$ & $\langle T^{\rm leaf}_{\rm dust}\rangle$ & $M_{\rm leaf}$ & $M_{\rm enc}$ & $\mathcal{R}$ \\
 & [ICRS] & [ICRS] & [$\mathrm{Jy}$] & [\arcsec] & [\arcsec] & [$\mathrm{deg}$] & [\arcsec] & [$\mathrm{pc}$] & [\arcsec] & [$\mathrm{pc}$] & [K] & [$M_\odot$] & [$M_\odot$] & \\
 \hline
1 & 14:45:20.98 & -59:49:30.4 & 3.513 & 28.0 & 9.9 & -159 & 16.7 & 0.23 & 16.1 & 0.22 & 23.6 & 485 & 369 & E \\
2 & 14:45:26.26 & -59:49:18.2 & 1.211 & 10.8 & 7.6 & 127 & 9.1 & 0.12 & 7.5 & 0.10 & 29.2 & 128 & 54 & E \\
3 & 14:44:56.43 & -59:47:57.2 & 0.845 & 20.1 & 11.0 & 58 & 14.9 & 0.20 & 14.2 & 0.19 & 18.5 & 159 & 244 & I \\
4 & 14:45:29.40 & -59:49:06.2 & 0.317 & 12.2 & 8.4 & 76 & 10.1 & 0.14 & 8.8 & 0.12 & 22.6 & 46 & 127 & E \\
5 & 14:45:12.26 & -59:49:00.2 & 0.293 & 13.6 & 8.0 & -140 & 10.4 & 0.14 & 9.1 & 0.12 & 20.5 & 48 & 53 & I \\
6 & 14:45:28.41 & -59:49:49.1 & 0.186 & 20.9 & 13.4 & 81 & 16.8 & 0.23 & 16.2 & 0.22 & 22.8 & 26 & 86 & E \\
7 & 14:45:00.00 & -59:48:34.8 & 0.180 & 6.9 & 5.0 & -152 & 5.9 & 0.08 & 2.6 & 0.04 & 19.2 & 32 & 37 & I \\
8 & 14:45:17.26 & -59:49:13.0 & 0.175 & 6.4 & 4.1 & -172 & 5.1 & 0.07 & - & - & 20.1 & 29 & 29 & E \\
9 & 14:44:59.85 & -59:48:55.9 & 0.126 & 6.9 & 4.6 & 153 & 5.6 & 0.08 & 1.8 & 0.03 & 22.5 & 18 & 13 & I \\
10 & 14:45:02.07 & -59:48:37.7 & 0.119 & 6.5 & 4.3 & -159 & 5.3 & 0.07 & - & - & 20.0 & 20 & 22 & I \\
\hline
\end{tabular}
\tablefoot{The ID is sorted in descending order of flux density $F_\nu$. $\theta_{\rm maj}$, $\theta_{\rm min}$: Measured major and minor FWHM. PA: Position angle of the elliptical model. $R_\theta$: Geometric mean FWHM in angular scale. $R_p$: Geometric mean FWHM in physical scale. $R_{\theta,\rm\ dec}$: Deconvolved geometric mean FWHM in angular scale. $R_{p,\rm\ dec}$: Deconvolved geometric mean FWHM in physical scale. $\langle T^{\rm leaf}_{\rm dust}\rangle$: Mean dust temperature of the leaf.
$M_{\rm leaf}$: Dendrogram leaf mass calculated from monochromatic flux. $M_{\rm enc}$: Dendrogram leaf calculated from enclosed pixels. $\mathcal{R}$: Region flag (Y for young, I for intermediate, and E for evolved). Only the first ten rows are shown; the complete version will be published in CDS.}
\end{table*}

The diagonalization of the second moments of the leaf structure determines the major and minor axis FWHM ($\theta_{\rm maj}$ and $\theta_{\rm min}$) as well as the PA. 
We calculated the geometric mean by $R_\theta = \sqrt{\theta_{\rm maj}\theta_{\rm min}}$. The deconvolved radius was then calculated as
\begin{equation}
    R_{\theta,\rm dec} = \eta\left[\left(\sigma_{\rm maj}^2-\sigma_{\rm bm}^2\right) \left(\sigma_{\rm min}^2-\sigma_{\rm bm}^2\right) \right]^{1/4}.
\end{equation}
Here $\sigma_{\rm maj}$ and $\sigma_{\rm min}$ are the dispersions of major and minor axis; $\sigma_{\rm bm}$ is the dispersion of radio beam $\sqrt{\theta_{\rm bmaj}\theta_{\rm bmin}/8\ln2}$; and $\eta$ is a factor that relates the rms size of the emission distribution to the angular radius of the object determined. The real $\eta$ depends on the emission distribution and its size relative to the beam. We adopted $\eta=2.4$, which is the median value derived in numerical simulations \citep{Rosolowsky2010}. The angular size was then converted to physical size by $R_{p,\rm dec} = R_{\theta,\rm dec}\, d$, where $d=2.8$~kpc is the distance of G316.8. For unresolved sources, deconvolution is invalid, and no values are given in the $R_{\theta,\rm dec}$ and $R_{p,\rm dec}$ columns. We calculated dense structure mass with monochromatic 1.3~mm flux density using the formula
\begin{equation}
    M_{\rm leaf} = \frac{F_\nu\, d^2}{\kappa_\nu B_\nu (\langle T^{\rm midres}_{\rm dust}\rangle)},
\end{equation}
where $\langle T^{\rm midres}_{\rm dust}\rangle$ is the mean dust temperature within the leaf mask from the 18\arcsec-resolution temperature map. We also calculated the enclosed mass $M_{\rm enc}$ within each leaf using Eq.~(\ref{eq:enc}). Throughout our work, we used $M_{\rm leaf}$ rather than $M_{\rm enc}$, as it measures the dense part of the hierarchical structures without including fluxes from the trunks. The leaves identified in this work have deconvolved physical sizes ranging from 0.02--0.2~pc, intermediate between the characteristic scales of cores and clumps \citep{Motte2018, Beuther2025}. Similar overdense structures have been reported in recent ACA 7m surveys of both infrared dark clouds \citep{Morii2024} and more evolved massive star-forming regions \citep{Xu2024Quarks2}. High-resolution follow-up observations in these studies reveal that such structures commonly host clustered internal substructures, namely groups of dense cores. These objects are therefore often referred to as sub-clump fragments (fragments hereafter) and are interpreted as potential progenitors of sub-clusters formed through non-unicentral, hierarchical collapse.

\subsection{Column density probability distribution function (N-PDF)} \label{result:npdf}

An important tool for characterizing interstellar medium or molecular clouds is N-PDFs. From theories of star-formation \citep{Padoan1997, VS2001, Hennebelle2008, Federrath2012, Burkhart2018}, the lognormal part can be attributed to isothermal supersonic turbulence \citep[e.g.,][]{Klessen2000, Federrath2013, Kortgen2019, Liu2025HVC}. One or more power-law tails (PLTs) can then develop under self-gravity \citep[e.g.,][]{Kritsuk2011, Girichidis2014, Jaupart2020, Donkov2021}. These theoretical studies and simulations include or exclude particular physical processes (such as solenoidal or compressive turbulence, stellar feedback, gravity, and magnetic fields) and thus demonstrate that the shape of an N-PDF strongly depends on the dominant process and the evolutionary state of the cloud. 
Observationally large variations in N-PDF shapes have been confirmed \citep[e.g.,][]{Lombardi2008, Kainulainen2009, Russeil2013, Stutz2015, Schneider2015a, Lin2016, Lin2017, Lin2022, Schneider2022}. For star-forming regions, at least one PLT was found with a slope between $-1.5$ and $-2$, consistent with self-gravity. A second PLT can be steeper or flatter, which is proposed to result from magnetic fields or stellar feedback. Recent work also examines whether a common, physically founded transition point exists between the lognormal and PLT parts of the N-PDFs. For example, \citet{Konyves2015} argued for a fixed value of around $A_v$ of 7--8, while \citet{Schneider2022} found no global transition point from their large sample of clouds from diffuse to high-mass star-forming regions. Other authors \citep{Alves2017} have even proposed that the N-PDF lacks a lognormal component and thus has no transition point.

With new high-resolution column density maps, we can conduct regional analyses to reveal local differences and their relation to evolutionary stages in the G316.8 filament. We define N-PDF as the probability of finding gas within a range [$\eta$, $\eta + \mathrm{d}\eta$], where $\eta$ is the natural logarithmic column density normalized by the mean column density $\eta \equiv \ln \left(N_{\rm H_2} / \langle N_{\rm H_2} \rangle \right)$. 
Column density maps are affected by line-of-sight (LOS) confusion, in particular in the Galactic plane and along spiral arms. In brief, unrelated molecular hydrogen along the LOS medium can lead to the overestimation of column density \citep{Schneider2022}. So, we adopted the same method as introduced in \citet{Schneider2015a} to determine the background and foreground contribution in a $5'\times5'$ box outside the bulk emission of the target. The $5'\times5'$ for our target is optimal because it is large enough to encompass the target but small enough to include selection bias. We removed a uniform $N^{\rm bkg}_{\rm H_2} = 7 \times 10^{21}$~cm$^{-2}$ throughout the image. The reliability of this method was verified in \citet{Schneider2015a, Ossenkopf2016}. For all other clouds, which are mostly molecular, we transformed H$_2$ column density into visual extinction, using the conversion formula $N_{\rm H_2} = A_v \times 0.94 \times 10^{21} \mathrm{cm}^{-2} \mathrm{mag}^{-1}$ \citep{Bohlin1978, Guver2009}.

\begin{figure}
\centering
\includegraphics[width=\linewidth]{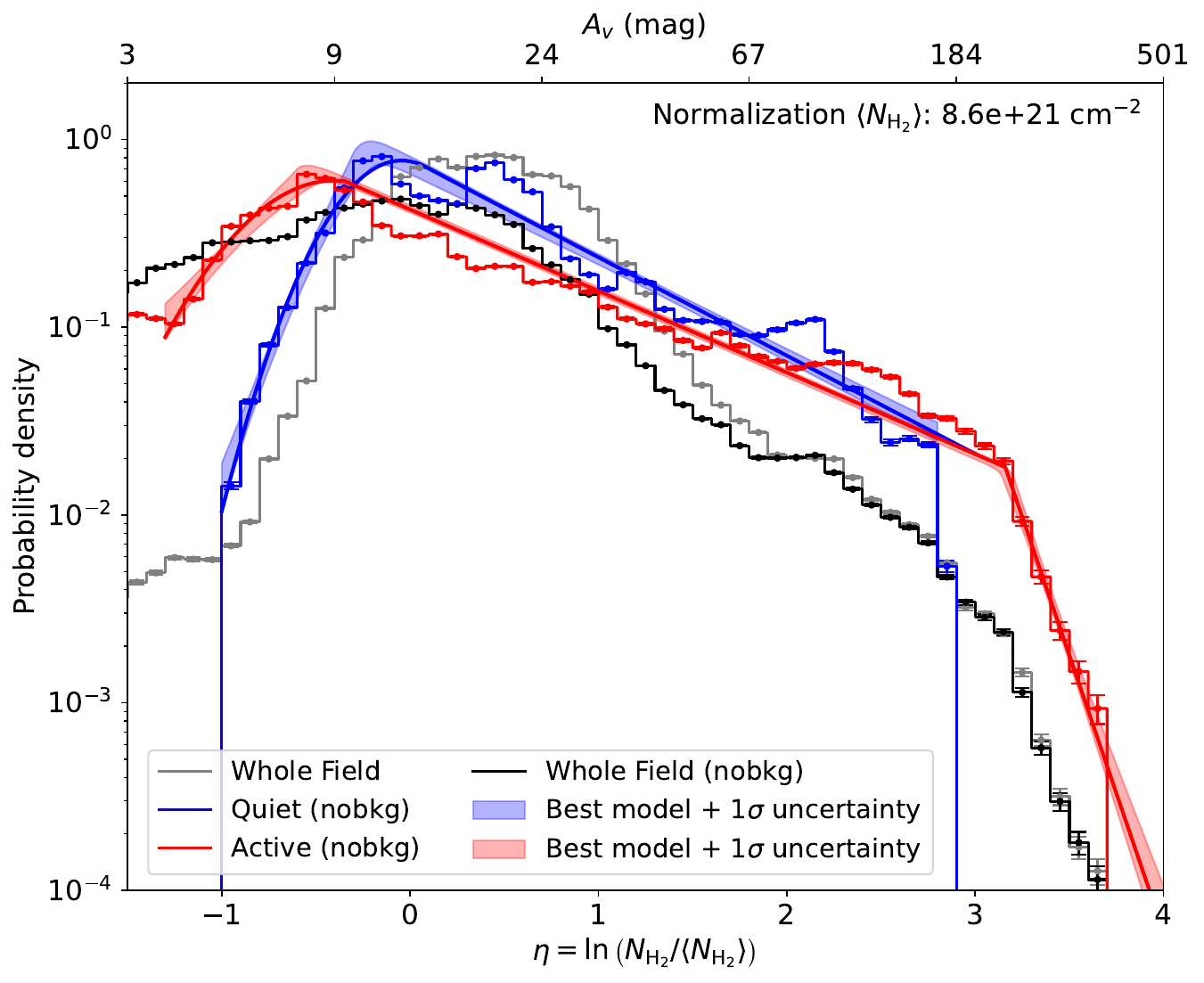}
\caption{N-PDFs for the entire region before and after background subtraction shown in gray and black. N-PDFs for the northern quiet and southern active parts are shown in blue and red. The best-fitting models and their posterior $1\sigma$ confidence intervals are shown with lines and shading. Visual extinction is also shown on the top axis.}
\label{fig:npdf}
\end{figure}

To construct comparative N-PDFs, we separated the G316.8 filament into the quiet and active regions as \citet{Watkins2019}. Specifically, the quiet region consists of the entire young part and the northern half of the intermediate part, while the active region consists of the entire evolved part and the southern half of the intermediate. As suggested by \citet{Schneider2015a}, the bin size of 0.1~dex provides the best compromise between resolving small features in the N-PDF and avoiding low pixel statistics. The N-PDFs over the entire G316.8 field and within the quiet and active parts of G316.8 are shown in Fig.~\ref{fig:npdf}. The N-PDF can be described by a lognormal in the low-density regime and one or two PLTs in the high-density regime. As shown in the comparison between black and gray N-PDFs, foreground and/or background contamination can have a strong narrowing effects on the lognormal part, which requires deduction \citep[e.g.,][]{Lim2016, Ossenkopf2016}. The N-PDFs in blue and red are both foreground- and background-subtracted. 

To understand the physical nature of N-PDFs, we performed model fitting of a lognormal part and one or two PLTs following the formula
\begin{equation} \label{eq:npdf}
p_\eta (\eta) =
\begin{cases}
\frac{1}{\sqrt{2\pi\sigma_\eta^2}} \exp\left[-\frac{(\eta - \mu)^2}{2\sigma_\eta^2}\right], & \eta \le b_1, \\[10pt]
C_1 \exp\left[-s_1 (\eta - b_1)\right], & b_1 < \eta \le b_2, \\[6pt]
C_2 \exp\left[-s_2 (\eta - b_2)\right], & \eta > b_2.
\end{cases}
\end{equation}
The free parameters $\mu$ and $\sigma_\eta$ describe the mean and dispersion of the normal distribution; $b_1$ and $b_2$ are the break points; and $s_1$ and $s_2$ are free parameters describing the PLTs. In general, two PLTs were used. Due to the continuity at the break points, $C_1$ and $C_2$ were determined, and six free parameters were fit. If only one PLT was assumed, $b_2$ was set as a high-density cut, and only four free parameters were fit. 

\begin{table*}
\caption{Parameterization of the regional column density map.}
\label{tab:measure}
\centering
\begin{tabular}{lcccccccc}
\hline
\hline
\\[-6pt]
\multirow{2}{*}{Subregion} & \multicolumn{6}{c}{N-PDF Fitting: lognormal + PLT(s)} & \multicolumn{2}{c}{$\Delta$-variance} \\
\cmidrule(lr){2-7} \cmidrule(lr){8-9}
 & $\mu$ & $\sigma$ & $b_1$ & $s_1$ & $b_2$ & $s_2$ & $\alpha_1$ & $\alpha_2$ \\
\hline
\multirow{2}{*}{young}
& \multirow{3}{*}{$1.16^{+2.38}_{-1.13}$}
& \multirow{3}{*}{$0.59^{+0.28}_{-0.25}$}
& \multirow{3}{*}{$-0.20^{+0.18}_{-0.07}$}
& \multirow{3}{*}{$-1.20^{+0.08}_{-0.08}$}
& \multirow{3}{*}{2.80}
& \multirow{3}{*}{-}
& \multirow{2}{*}{2.7(0.2)}
& \multirow{2}{*}{2.3(0.2)}
\\
& & & & & & & & \\ 
\multirow{2}{*}{intermediate}
& & & & & &
& \multirow{2}{*}{1.9(0.1)}
& \multirow{2}{*}{2.1(0.2)}
\\
& \multirow{3}{*}{$2.30^{+2.35}_{-2.39}$}
& \multirow{3}{*}{$1.17^{+0.37}_{-0.54}$}
& \multirow{3}{*}{$-0.50^{+0.10}_{-0.07}$}
& \multirow{3}{*}{$-1.00^{+0.01}_{-0.01}$}
& \multirow{3}{*}{$3.15^{+0.03}_{-0.04}$}
& \multirow{3}{*}{$-7.18^{+0.76}_{-0.72}$}
& & \\
\multirow{2}{*}{evolved}
& & & & & & 
& \multirow{2}{*}{1.5(0.1)} 
& \multirow{2}{*}{1.4(0.0)} 
\\
& & & & & & & \\
\hline
\end{tabular}
\tablefoot{Columns 2--7 list the MCMC fitting parameters for the quiet (young$+$$\frac{1}{2}$intermediate) and active ($\frac{1}{2}$intermediate$+$evolved) regions. The quoted values denote the median (50th percentile) with the 16th and 84th percentiles. Columns 8--9 present the two power-law slopes of $\Delta$-variance for the three subregions in $\alpha_1$ and $\alpha_2$.}
\end{table*}

Fitting was performed using the Markov chain Monte Carlo (MCMC) method, which naturally accounts for parameter covariances and provides robust uncertainty estimates for posteriors. More fitting details can be found in Appendix~\ref{app:mcmc_fit}. The low-extinction regime ($<-1<\eta<0$) can be modeled well with one lognormal distribution. The dispersion of the active part is twice broader than that of quiet part, which aligns with more turbulent environments in the active star-forming regions. At the high-density regime, the south N-PDF is well described by two PLTs. The first one has a slope of about $-1.0$ and steepens to about $-7.2$ at the break point of $\eta=3.14$ or $2\times10^{23}$~cm$^{-2}$. However, the northern quiet part is better described by a one-PLT model with a hard cutoff at $\eta=2.80$ or $1\times10^{23}$~cm$^{-2}$ and a slightly steeper PLT slope of $-1.2$ than in the active part. All the best-fit parameters are summarized in Table~\ref{tab:measure}, and the models as well as their $1\sigma$ confidence levels are shown by the shaded regions in Fig.~\ref{fig:npdf}.

\subsection{Structural analyses $\Delta$-variance} \label{result:delta}

We applied the $\Delta$-variance technique to the multi-resolution H$_2$ column-density map. The $\Delta$-variance method \citep{Stutzki1998, Ossenkopf2008a, Ossenkopf2008b} allows us to measure structure variations in a map $S$ on a given scale $\ell$ (called ``lag'') by filtering the map with a circularly symmetric wavelet $\bigcirc_L$:
\begin{equation}
    \sigma_\Delta^2 (\ell) = \langle (S \otimes \bigcirc_\ell)^2 \rangle_{x,y}.
\end{equation}
In the formula, $\langle\ldots\rangle_{x,y}$ is the ensemble average over coordinates $x$ and $y$ in the column density map, and $\otimes$ is the convolution operator. We used a filter function with an annulus-to-core-diameter ratio of about 1.5 since it provides the best results for a clear detection of pronounced scales \citep{Ossenkopf2008a, Ossenkopf2008b}. Point-wise uncertainties on $\sigma_{\Delta}^{2}(\ell)$ were obtained via block bootstrap. The region was partitioned into $n_{\rm tile}$ equal tiles, resampled with replacement, and the full pipeline was rerun over 500 realizations to capture sampling variance and edge effects. Fifty lags were sampled from 0.04 and 2.04~pc.

\begin{figure}[!ht]
    \centering
    \includegraphics[width=\linewidth]{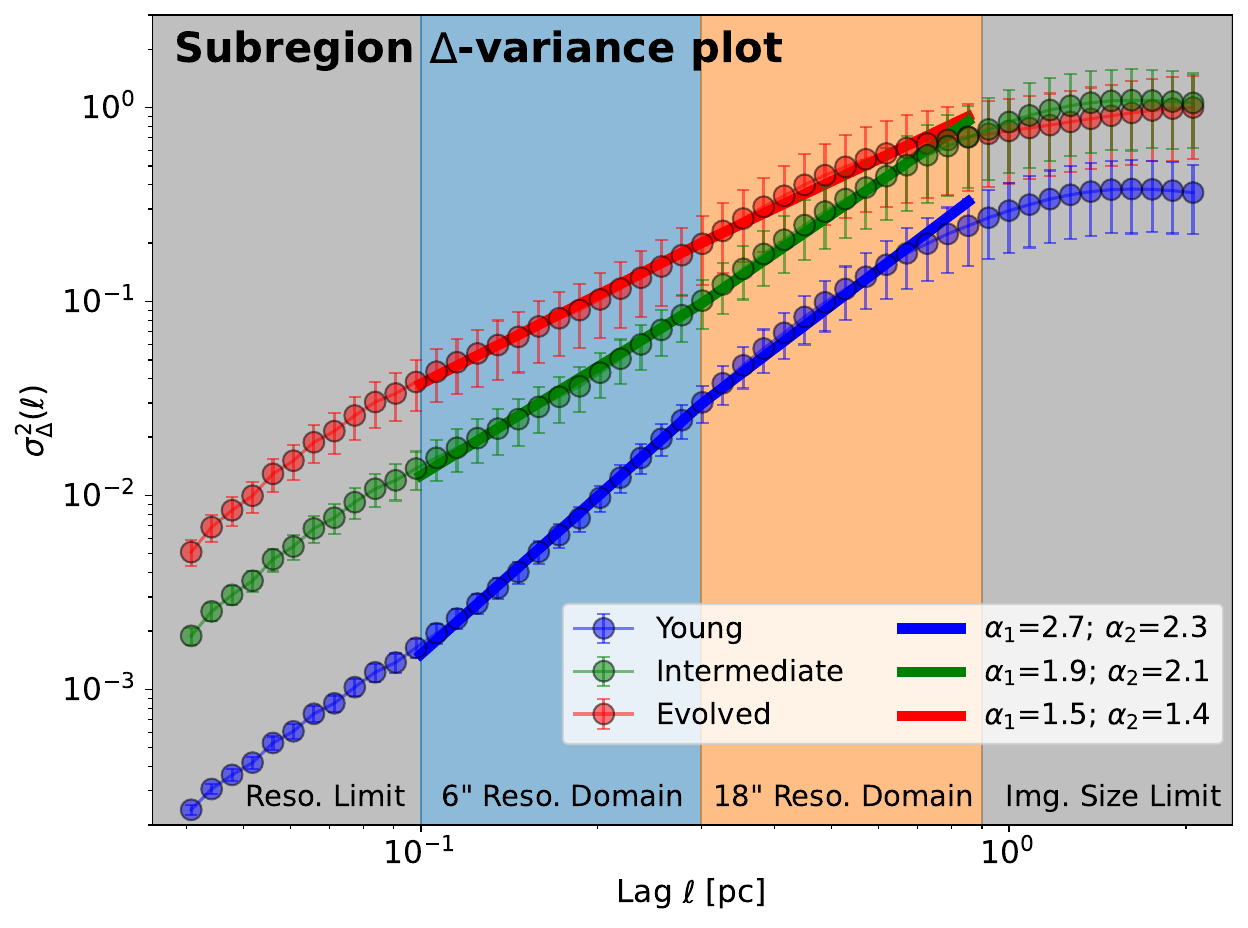}
    \caption{$\Delta$-variance spectrum for the three subregions. The blue, green, and red dots represent the young, intermediate, and evolved subregions, respectively. High- and low-resolution domains span 0.1 to 0.3~pc (blue shading) and 0.3 to 0.9~pc (orange shading), respectively. The left and right gray-shaded regions indicate the resolution-limit and image-size-limit domains, respectively. The broken power-law fitting results are shown with solid colored lines.}
    \label{fig:delta-variance}
\end{figure}

The $\Delta$-variance spectra of the three subregions are shown with colored curves in Fig.~\ref{fig:delta-variance}. Overall, each follows a power-law form. Because the angular resolution is nonuniform, $\sigma_{\Delta}^{2}$ at scales $\ell<18\arcsec$ is dominated by the high-resolution footprint, whereas $\ell>18\arcsec$ increasingly samples the lower-resolution component. This naturally produces a break in slope near the coarse-beam scale, i.e., at $\ell\simeq18\arcsec$ ($\approx0.25$~pc at 2.8~kpc), marking the transition between the high- and low-resolution domains. The curves also show a mild steepening at the smallest lags and clear flattening at the largest lags. The former arises from enhanced correlations due to beam over-sampling \citep{Bensch2001}, while the latter is primarily a finite-map (window) effect. Accordingly, we fit a double power-law (DPL) model over $\ell\in[0.1,\,0.9]$~pc with a fixed transition at 0.3~pc, excluding the beam-dominated and map-size–affected ends of the spectrum. 

We determined $\alpha$ through linear regression weighted by $1\sigma$ errors from the bootstrap distribution. The fitting results are listed in Cols.~8--9 of Table~\ref{tab:measure}. For the evolved subregion, the DPL has a small-scale slope of 1.5 and a large-scale slope of 1.4. For the intermediate subregion, they are as steep as 1.9 and 2.1. For the young subregion, they are even steeper at 2.7 and 2.3. Their uncertainties are typically 10\%. 

\section{Regional differences capture dense gas evolution} 
\label{sec:evolution}

\subsection{Dense fragments and structural hierarchy} \label{evolution:fragment}

The dendrograms in Fig.~\ref{fig:astrodendro} visualize the hierarchy of dense structures and its difference across subregions. Trees that exhibit multilevel branching host leaves exclusively from the intermediate and evolved subregions, whereas leaves from the young subregion occur only in isolated, nonhierarchical trees. This segregation suggests that structural hierarchy emerges and likely strengthens as star formation progresses. 

\begin{figure}[!ht]
\centering
\includegraphics[width=\linewidth]{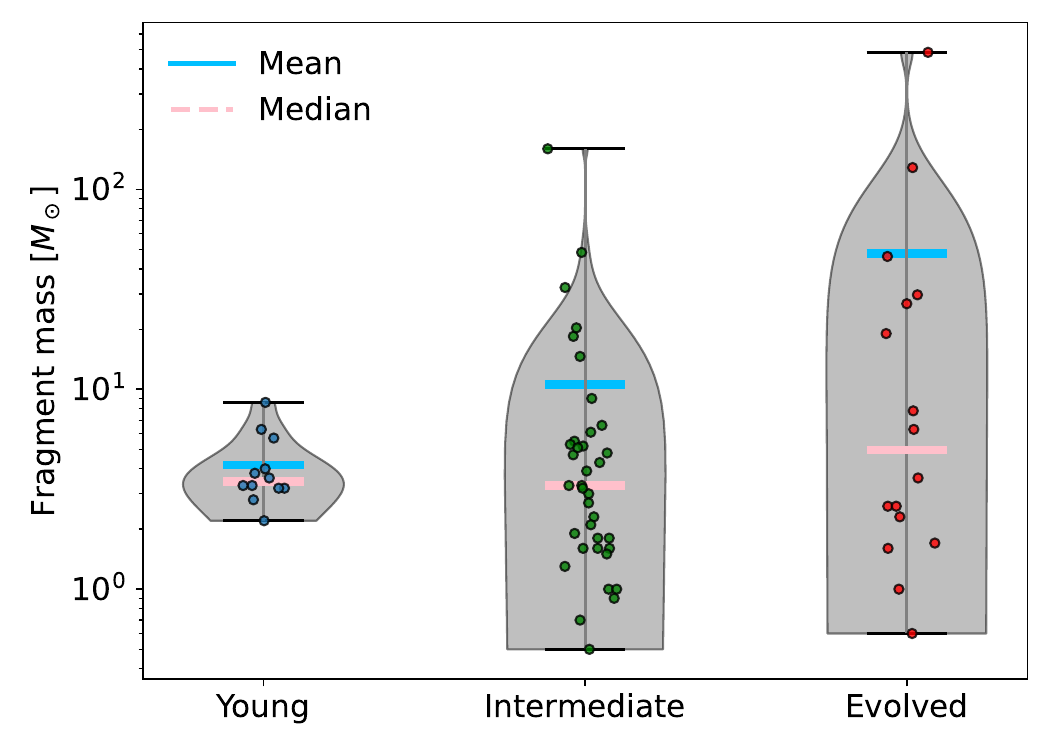}
\caption{Comparison of fragment masses among the young, intermediate, and evolved subregions. The violin plot shows the extrema (black dashes), medians (pink), and means (blue).}
\label{fig:fragmass}
\end{figure}
Figure~\ref{fig:fragmass} compares the fragment-mass distributions across the three subregions. From the young, intermediate, and evolved stages, the maximum fragment mass increases from $\sim$8 to 160 and reaches $490~M_\odot$, indicating substantial mass accumulation into dense structures as evolution proceeds. Even accounting for possible free-free contamination, the maximum fragment mass in the evolved region would be overestimated by at most $\sim$14\% (see discussion in Sect.~\ref{result:sed}) and thus remains significantly higher than in the other two subregions. This progressive mass growth is consistent with recent interferometric surveys, which similarly report increasingly massive cores in more evolved environments \citep[e.g.,][]{Traficante2023, Pouteau2023, Xu2024Assemble1, Xu2024Quarks2, ALMAGAL3, Kinman2025}.

In contrast, the apparent decrease in the minimum fragment mass from 2.2 to $0.6~M_\odot$ is unlikely to reflect intrinsic physical evolution but rather arises from observational biases. The continuum mass sensitivity depends on beam size, noise level, and dust temperature, as $\sigma_{\rm mass}\propto d^2\,\sigma_{\rm rms}\,B_\nu(T)^{-1}\,\Omega_{\rm beam}$. Because the southern regions are warmer and brighter, their mass sensitivities are significantly improved. For instance, adopting characteristic dust temperatures of 10~K in the young region and 30~K in the evolved region yields mass sensitivities of $\sim\!0.9$ and $\sim\!0.2$~\msun, respectively. Low-mass fragments remain undetected in the colder subregions owing to limited sensitivity.

The more physically meaningful evolutionary signature is therefore the increasing dynamic range of fragment masses, as reflected by the progressively broader mass distributions in Fig.~\ref{fig:fragmass}. Such broadening is also observed in recent ALMA surveys \citep[e.g.,][]{Morii2024, ALMAGAL3} and likely reflects increasingly nonuniform mass accretion among dense structures as evolution proceeds. Summing the fragment masses within each subregion yields total dense gas masses of 50, 391, and $766~M_\odot$ in the young, intermediate, and evolved regions, respectively. Defining regional dense gas fraction as $f_{\rm dg}(\mathcal{R}) \equiv \sum M_{\rm leaf}/M_{\rm cloud}(\mathcal{R})$, we obtain $0.4\%$, $2.3\%$, and $9.6\%$, respectively. This monotonic increase indicates that an increasingly high fraction of the cloud mass becomes locked into compact structures as evolution proceeds, consistent with ACA 7~m surveys of massive clumps that find $f_{\rm dg}$ rising from $<1$\% to $\sim$10\% \citep{Xu2024Quarks2}. A similar behavior has been reported by \citet{Diaz2023}, who find that the leaf mass efficiency increases rapidly with molecular gas column density and interpret this trend as a signature of enhanced star formation activity.

\subsection{Column density probability distribution functions} \label{evolution:npdf}

The higher fraction of dense structures in the more active region is consistent with the trends revealed by the N-PDFs. As shown in Fig.~\ref{fig:npdf}, the flatter primary PLT and the emergence of a second PLT both indicate that, in active regions, a significant fraction of the gas resides in high-density structures. However, the primary PLT in the active region is too flat to be explained solely by self-gravity. Additional mechanisms, such as stellar radiative feedback or other processes that slow down global collapse and reduce the mass flow toward higher densities, are likely at play \citep{Tremblin2014, Schneider2015b}. Indeed, as seen in Fig.~\ref{fig:observe}, the active region hosts an evolved bipolar H{\sc ii} region that originates directly from the main filament ridge, providing strong evidence that feedback-driven compression reshapes the dense-gas distribution. As discussed by \citet{Watkins2019}, although the majority of the ridge remains gravitationally bound ($\alpha_{\rm vir}\le2$), small gas pockets near high-mass stars may already be unbound by feedback. Further molecular-line analyses are therefore essential for constraining the detailed gas dynamics. Comparative studies of other multistage filaments, such as W43-MM2\&MM3, have likewise shown that post-(star)burst clouds exhibit flatter N-PDFs than their pre-burst counterparts \citep{Pouteau2023}. These results collectively suggest that the cloud's internal structure is tightly linked to its evolutionary stage, even though the causality between two remains undetermined. For the second, steeper slope in the active region, one possibility is an orientation perpendicular to the LOS column density distribution, as proposed by \citet{Schneider2022}. However, no firm conclusion on this subject exists, and simulations still struggle to correctly reproduce these flatter or steeper second PLTs. 

\subsection{$\Delta$-variance} \label{evolution:delta}

An image can be formulated as $P(k)\propto k^{-\beta}$, where $P(k)$ represents the power spectral density versus wave-number magnitude $k=2\pi/\lambda$, which is inverse to spatial scale. Its $\Delta$-variance scales as $\sigma_\Delta^2\propto \ell^{\alpha}$ with $\alpha=\beta-2$. Shallower $\alpha$ therefore implies relatively more small-scale power. In G316.8, we find: i) The evolved and intermediate subregions have similar slopes across scales, while the young region has a much steeper slope on smaller scales. ii) The slope seems to systematically flatten with age: steepest for the young, flatter for the intermediate, and flattest for the evolved. These findings suggest that molecular cloud density structures vary across different evolutionary stages. In the early stage, clouds contain far fewer dense structures at small scales. Consequently, the $\Delta$-variance changes rapidly at small lags, yielding large $\alpha_1$. When the clouds further evolve, dense structures accumulate due to gravity or gas accretion flows, increasing the density contrast. The $\Delta$-variance changes much more slowly, yielding smaller $\alpha_1$. However, the $\alpha_2$ difference between the young and evolved subregions is not as large as $\alpha_1$. This might indicate that the large-scale structures have longer evolution timescales, as formulated by $t_{\rm ff} \sim \rho^{-1/2} \sim r^{3/4}$ for a density profile $\rho \sim r^{-1.5}$.

\subsection{Potential bias and caveats} \label{evolution:bias}

It should be noted that the chosen dust-opacity model can dominate the systematic uncertainties in SED fitting, exceeding the formal flux calibration uncertainties discussed in Sect.~\ref{result:sed}. To assess this effect, we refit the data using two widely adopted prescriptions: the cloud-scale opacity of H83 and the protostellar core opacities of \citet[][OH94]{Ossenkopf1994}. For the OH94 model, we adopted grains with ice mantles at a density of $n(\mathrm{H}_2)=10^5~\mathrm{cm^{-3}}$, consistent with the highest densities inferred among the three subregions. Relative to H83, the OH94 assumption yields lower column densities of $27\pm1$\% and higher dust temperatures higher of $9\pm1$\%. This implies that the absolute values of $N_{\rm H_2}$ and $T_{\rm dust}$ are uncertain at the $\sim$30\% level or higher. However, for a given dust opacity prescription, the resulting offsets are largely coherent across the subregions, such that the relative variations in $N_{\rm H_2}$ and $T_{\rm dust}$ are more stable. Consequently, while absolute column densities and temperatures should be interpreted with caution, the relative differences between the subregions and the evolutionary trends inferred in this work remain robust. It is also important to note that our evolutionary assessments primarily concerns the cold, high-density molecular gas in the G316.8 region, while gas phases most directly affected by stellar feedback are not explicitly traced. Consistently, the dense gas observed in G316.8 appears only weakly heated by the H{\sc ii} regions (see Fig.~\ref{fig:multires}) and shows no clear signs of large-scale disruption. 

\section{Conclusions and perspectives} \label{sec:conclude}

We presented a 118-pointing mosaic of the 14-parsec-long linear filament G316.8, which exhibits a clear in situ evolutionary sequence from the northern (young), central (intermediate), and southern (evolved) subregions. In this first paper of the series, we processed and analyzed the Atacama Compact Array (ACA) 7m continuum data and combined them with \textit{Herschel} and APEX data as zero spacings. We produced multi-resolution temperature and column-density images with effective angular resolutions of $18\arcsec$ (global) and $6\arcsec$ (over the highest–density footprint), corresponding to $\sim$0.24~pc and $\sim$0.08~pc at 2.8~kpc. 

Using the high-resolution column density image, we performed subregion analyses in three ways: (i) dense-fragment statistics, (ii) column-density probability distribution functions (N-PDFs), and (iii) $\Delta$-variance. All three diagnostics reveal clear subregional differences. Along the sequence young to intermediate to evolved, the maximum fragment mass and the regional dense-gas mass fraction both increase, indicating progressive concentration of mass into compact (sub-parsec) structures. The N-PDFs evolve from a sharp cutoff to a flatter first PLT with a steep secondary PLT at the highest columns. The $\Delta$-variance spectra progressively show shallower slopes (more small-scale contrast) in the more evolved regions. Taken together, these results lead to a consistent evolutionary picture: as filamentary cloud evolves, molecular gas is funneled into increasingly dense and structured sub-parsec components. 

This evolutionary framework provides a foundation for future work linking these structural measurements to gas kinematics and dense core formation. In forthcoming papers, we will address three key aspects of massive star cluster formation. First, $\sim$800~AU resolution ALMA 12m data will enable a complete census of dense cores along the 14~pc filament, allowing the core mass function and spatial distribution to be interpreted in an evolutionary context and directly compared with recent simulations and synthetic ALMA observations \citep[e.g.,][]{JD2023, Hsu2023, Rosetta1, Rosetta2, Rosetta3}. Second, the spectral-line data, particularly C$^{18}$O and H$_2$CO, will provide a powerful probe of gas flows along the filament and into individual cores, offering a benchmark for testing fiber formation and accretion scenarios in numerical models \citep[e.g.,][]{Hacar2024, Grudic2021}. Finally, G316.8 constitutes a unique laboratory for investigating the evolving role of magnetic fields in massive star formation, a topic that will be further explored with newly approved ALMA polarization observations.

\section*{Data availability}
Table~\ref{tab:catalog} are available in electronic form at the CDS via anonymous ftp to cdsarc.u-strasbg.fr (130.79.128.5) or via http://cdsweb.u-strasbg.fr/cgi-bin/qcat?J/A+A/.

\begin{acknowledgements}

We thank the anonymous referee for improving and polishing the paper. This work has been supported by the National Natural Science Foundation of China (NSFC, Grant No. 12033005). F.W.X appreciates helpful discussions with Zhiyu Zhang at Xiamen, Hendrik Linz at MPIA Heidelberg, and with Siju Zhang and Yuxin He at Shanghai. F.W.X thanks Dr. Yue Li for providing working place and company during writing the paper in Shenzhen. N.S. acknowledges support from the CRC 1601 (SFB 1601 sub-project B2) funded by the DFG – 500700252.
RGM acknowledges the support of UNAM-PAPIIT project IN105225. G.G. gratefully acknowledges support by the ANID BASAL project FB210003. AG acknowledges support from the NSF under CAREER grant 2142300. H.B.L. is supported by the National Science and Technology Council (NSTC) of Taiwan (Grant Nos. 113-2112-M-110-022-MY3). This work makes use of \textit{turbstat} \citep{Kock2019}, \textit{astropy} \citep{Astropy2022}, \textit{scipy} \citep{Scipy2020}, and \textit{numpy} \citep{Numpy2020}. This paper makes use of the following ALMA data: ADS/JAO.ALMA\#2016.1.00909.S. ALMA is a partnership of ESO (representing its member states), NSF (USA) and NINS (Japan), together with NRC (Canada), NSTC and ASIAA (Taiwan), and KASI (Republic of Korea), in cooperation with the Republic of Chile. The Joint ALMA Observatory is operated by ESO, AUI/NRAO and NAOJ. In addition, publications from NA authors must include the standard NRAO acknowledgement: The National Radio Astronomy Observatory is a facility of the National Science Foundation operated under cooperative agreement by Associated Universities, Inc.

\end{acknowledgements}

\bibliographystyle{aa} 
\bibliography{g316_ref} 

\begin{appendix}

\section{Spectral energy distribution fitting} \label{app:sedfitting}

The SED was fit by assuming a single-component modified blackbody emission as
\begin{equation} \label{eq:mbb}
    I_\nu = B_\nu(T_\mathrm{d})(1-e^{-\tau_\nu}),
\end{equation}
where $B_\nu (T_\mathrm{d})$ is the blackbody emission at given dust temperature $T_\mathrm{d}$,
\begin{equation} \label{eq:bb}
    B_\nu (T_\mathrm{d}) = \frac{2h\nu^3}{c^2}\frac{1}{e^{h\nu/k_\mathrm{B}T_{d}}-1},
\end{equation}
and $\tau_\nu$ is the opacity at frequency $\nu$,
\begin{equation} \label{eq:tau}
    \tau_\nu = \mu m_{\mathrm{H}} \kappa_\nu N_{\text{H}_2},
\end{equation}
where $\mu$ is the mean molecular weight relative to the mass of hydrogen atom $m_\mathrm{H}$ and is taken as 2.8 assuming $n_{\rm He} = 0.1\,n_{\rm H}$ with ignoring other heavier elements. Wavelength-dependent dust opacity is given by
\begin{equation} \label{eq:kappa}
    \kappa_\lambda = \kappa_0 \left(\frac{\lambda}{\lambda_0}\right)^{-\beta},
\end{equation}
where we used the reference dust opacity per unit gas and dust mass $\kappa_0=0.1$~cm$^2$~g$^{-1}$ at the reference wavelength of 300~$\mu$m and dust opacity spectral index $\beta=2.0$ \citep{Hildebrand1983}. 

\begin{figure}[!ht]
    \centering
    \includegraphics[width=\linewidth]{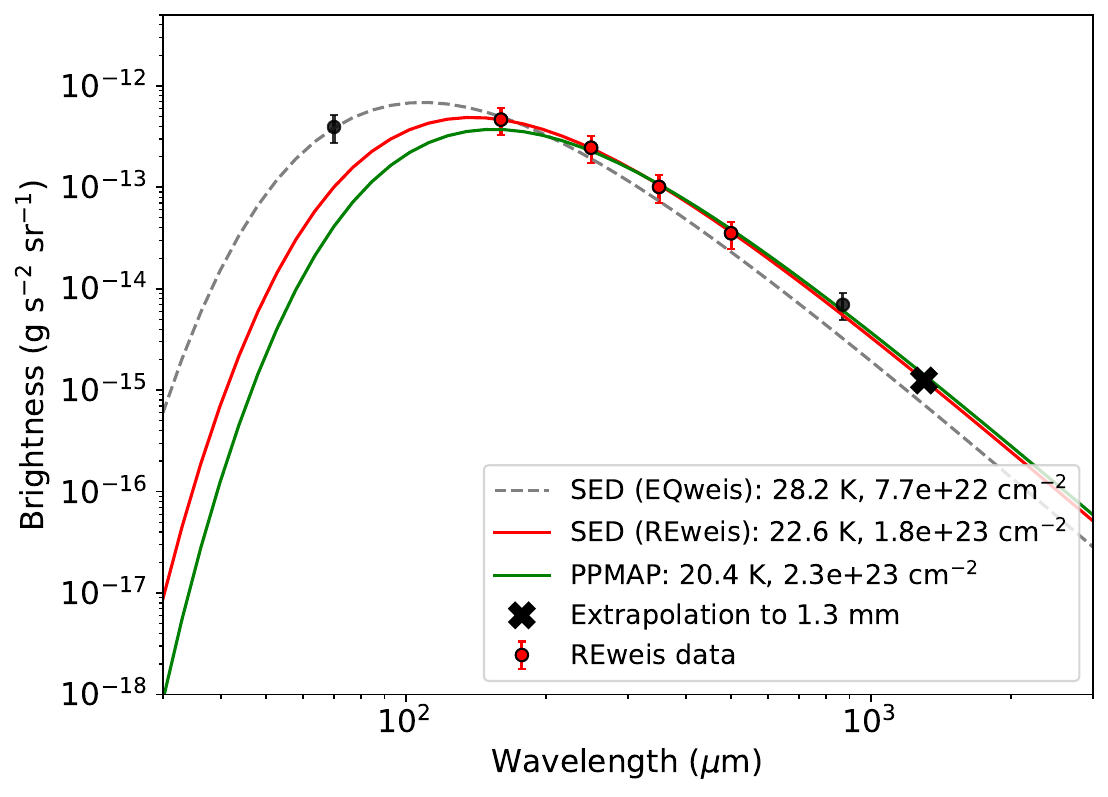}
    \caption{SED fitting of the brightest pixel. The gray-dashed line indicates the case where 70, 160, 250, 350, 500~$\mu$m, and 870~$\mu$m data have equal weights (EQweis), while the red solid line shows the reweighted case (Reweis) where the 70 and 870~$\mu$m data are discarded for fitting. The PPMAP result \citep{Marsh2017} is shown in green solid line for comparison.}
    \label{fig:sedpixel}
\end{figure}

In Fig.~\ref{fig:sedpixel}, we showcase the SED fitting in the brightest pixel of G316.8 region. The example demonstrates that 70~$\mu$m can be easily affected by a warm component especially toward protostellar objects. So REweis describes the long wavelengths ($\lambda > 160$~$\mu$m) much better than EQweis. This is essential especially to those active star-forming regions where the inclusion of the 70~$\mu$m data easily overestimate the temperature and underestimate the column density. The real case here is that REweis gives 22.9~K while EQweis gives 28.2~K, so the column density by REweis is a factor of 2 higher than the EQweis. Besides, the 870~$\mu$m point deviates significantly from the model which is supposed to be contaminated by the CO (3-2) line emission. With the Reweis model, we extrapolate the 18\farcs2 image to 1.3~mm for further image combination with interferometric data.

\section{Image combination} \label{app:comb}

\subsection{The combination in the image domain} \label{app:comb:xy}

To combine data at different resolutions ($I_{\rm hires}$ and $I_{\rm lores}$) in the image domain, we adopt a weighted blending scheme that preserves high-resolution detail wherever available while smoothly incorporating low-resolution information in uncovered regions. The procedure can be described as the following steps: 1) to up-sample the low-resolution to match the pixel dimensions of the high-resolution image; 2) to align the flux between two images by relative calibration; 3) to construct a weighting mask with values ranging from 1 in regions covered by high-resolution data to 0 in regions without such coverage. For each pixel, we compute the distance to the nearest missing pixel in the high-resolution image ($d_{\mathrm{out}}$) and the distance to the nearest valid pixel ($d_{\mathrm{in}}$). These distances are combined to give
\begin{equation}
    \alpha = \frac{d_{\mathrm{in}}}{d_{\mathrm{in}} + d_{\mathrm{out}} + \epsilon} ,
\end{equation}
where $\epsilon$ is a small constant to avoid division by zero. The mask is then smoothed with a Gaussian kernel (width set by the \texttt{weighting\_smooth} parameter) to avoid sharp seams.

The final merged image is obtained by weighted combination:
\begin{equation}
    I_{\mathrm{merged}} = \alpha \, I_{\mathrm{hires}} + (1 - \alpha) \, I^{\mathrm{calib}}_{\mathrm{lores}} ,
\end{equation}
where $I_{\mathrm{hires}}$ is the gap-filled high-resolution image and $I^{\mathrm{calib}}_{\mathrm{lores}}$ is the calibrated, resampled low-resolution image. Practically, $I_{\mathrm{merged}}$ preserves the information from both high and low resolution images. 

\subsection{The combination in th uv domain} \label{app:comb:uv}

\begin{figure*}[!ht]
    \centering
    \includegraphics[width=1.0\linewidth]{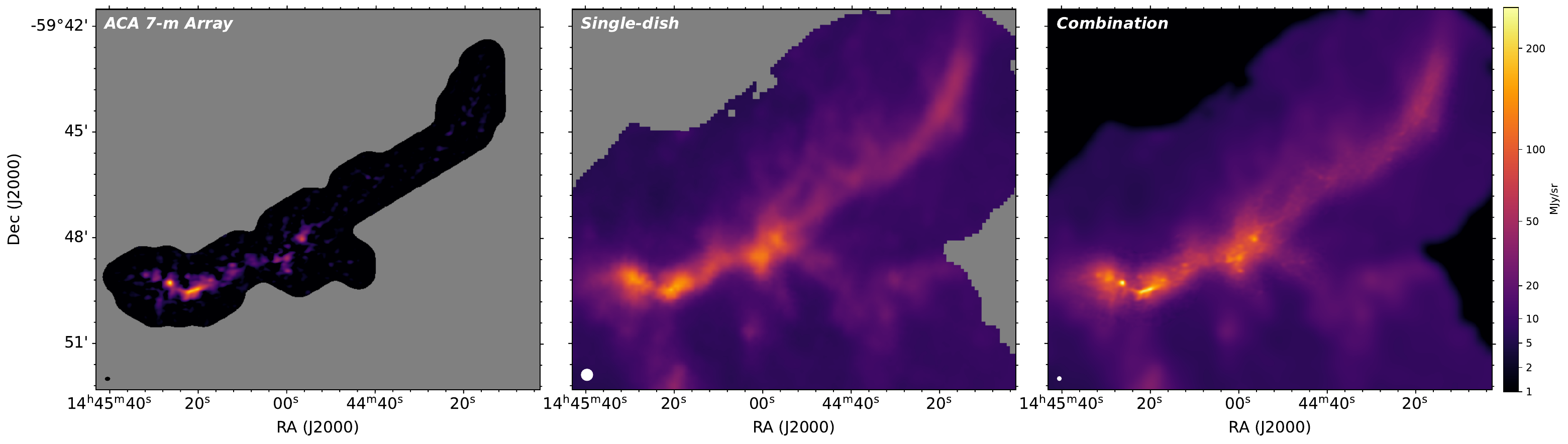}
    \caption{ACA 7m array (left), single dish (middle), and combined (right). The color scales are the same for all three panels. }
    \label{fig:imgcomb}
\end{figure*}

Interferometric (Interf) observations lose the short-spacing information which can be compensated by single-dish (SD) telescope. We adopt the J-comb linear combination algorithm \citep{Jiao2022} in the uv domain to combine the SD and Interf images. The uv range of Interf and SD data overlaps between 18\farcs2 (SD resolution) and 28\farcs0 (Interf MRS), so the combination is more robust. The relative flux calibration is performed using point source. As shown in Fig.~\ref{fig:imgcomb}, the Interf image shows substantial missing flux at large scale, while the poorer resolution of the SD image smooth out fine structures. The image combination takes advantage of two, achieving both high resolution and complete coverage of large-scale flux. More importantly, the interferometric `negative bowl' around the bright source in the original Interf image also disappears, increasing the image fidelity. 

\section{MCMC fitting of the N-PDF}
\label{app:mcmc_fit}

We model the N-PDF with functional forms given in Eq.~(\ref{eq:npdf}). For each bin center $\eta_i$ and measured $p_i(\eta)$ with uncertainty $\sigma_{i}(\eta)$, the likelihood function is given by
\begin{equation}
\ln \mathcal{L}(\boldsymbol{\theta}) =
- \frac{1}{2} \sum_{i=1}^n
\left[
\frac{p_i - p_{\mathrm{model}}(\eta_i \mid \boldsymbol{\theta})}{\sigma_i}
\right]^2
- \sum_{i=1}^n \ln \left( \sqrt{2\pi} \, \sigma_i \right),
\end{equation}
where $\boldsymbol{\theta}$ denotes the parameter set (e.g., $\mu, \sigma, b_1, s_1, C_1$ for the one-tail model). We adopt broad, uniform priors to ensure flexibility while avoiding nonphysical solutions. Besides, we include an intrinsic scatter $s_{\mathrm{int}}$ to avoid sharp likelihood function, though in our final fits we set $s_{\mathrm{int}} = 0$. We employ the Markov Chain Monte Carlo (MCMC) ensemble sampler implemented in the \texttt{emcee} Python package \citep{Foreman2013}. From the flattened posterior chain, we obtain the best-fit model and calculate 16th and 84th percentiles as 1$\sigma$ confidence region in the N-PDF plots. 

\end{appendix}

\end{document}